\documentclass[a4paper,11pt]{article}

\pdfoutput=1
\usepackage[utf8]{inputenc}
\usepackage{textcomp}
\usepackage{amsmath}
\usepackage{amssymb}
\usepackage{amsfonts}
\usepackage{mathrsfs}
\usepackage{graphicx}
\usepackage{booktabs,adjustbox}
\usepackage{caption}
\usepackage{slashed}
\usepackage{verbatim}
\usepackage{float}
\usepackage{setspace}
\usepackage{subfigure}
\usepackage{jheppub}

\usepackage{color}
\usepackage{ulem}

\usepackage{bookmark}
\usepackage{wasysym}

\makeatletter
\newcommand{\thickhline}{%
	\noalign {\ifnum 0=`}\fi \hrule height 1pt
	\futurelet \reserved@a \@xhline
}
\makeatother

\allowdisplaybreaks[1]
\begin{document} 
\title{\boldmath Linking the $R_{K^{(\ast)}}$ anomalies to the Hubble tension via a single right-handed neutrino}

\author[a]{Wen-Feng Duan,}
\author[a]{Shao-Ping Li,}
\author[a,1]{Xin-Qiang Li\note{Corresponding author.},}
\author[a,b]{and Ya-Dong Yang}

\affiliation[a]{ Institute of Particle Physics and Key Laboratory of Quark and Lepton Physics~(MOE), Central China Normal University, Wuhan, Hubei 430079, China}
\affiliation[b]{ Institute of Particle and Nuclear Physics, Henan Normal University, Xinxiang 453007, China}

\emailAdd{dufewe@mails.ccnu.edu.cn}
\emailAdd{ShowpingLee@mails.ccnu.edu.cn}
\emailAdd{xqli@mail.ccnu.edu.cn}
\emailAdd{yangyd@mail.ccnu.edu.cn}
	

\abstract{The updated measurements from the LHCb and SH0ES collaborations have respectively strengthened the deviations of the ratio $R_{K}$ in rare semi-leptonic $B$-meson decays and the present-day Hubble parameter $H_0$ in the Universe, implying tantalizing hints of new physics beyond the Standard Model. In this paper, we consider a simple flavor-specific two-Higgs-doublet model, where the long-standing $R_{K^{(*)}}$ anomalies can be addressed by a one-flavor right-handed neutrino. One of the intriguing predictions resulting from the parameter space for the $R_{K^{(*)}}$ resolution under flavor- and collider-physics constraints points toward a shift of the effective neutrino number, $\Delta N_{\rm eff}=N_{\rm  eff}-N_{\rm eff}^{\rm SM}$, as favored to ease the $H_0$ tension. Depending on whether the neutrino is of Dirac or of Majorana type, we show that the resulting shift is $\Delta N_{\rm eff}\simeq 1.0$ for the former and $\Delta N_{\rm eff}\simeq 0.5$ for the latter case, respectively. While the Dirac case is disfavored by the CMB polarization measurements, the Majorana solution is consistent with the recent studies via a combined data set from various sources. Consequently, such a simple flavor-specific two-Higgs-doublet model provides a link between the $R_{K^{(*)}}$ anomalies and the $H_0$ tension, which in turn can be readily verified or falsified by the upcoming measurements.}

\maketitle

\section{Introduction}
\label{intro}
In rare semi-leptonic $B$-meson decays, there exist a series of long-standing deviations between the Standard Model (SM) predictions and the LHCb measurements~\cite{LHCb:2013ghj,LHCb:2014vgu,LHCb:2017avl,LHCb:2019hip,LHCb:2021trn,LHCb:2021lvy}. In particular, the ratios $R_{K^{(\ast)}}$, which are defined by 
\begin{align} \label{RK}
	R_{K^{(\ast)}} \equiv \frac{\mathcal{B} (B \to K^{(\ast)} \mu^+ \mu^-)}{\mathcal{B} (B \to K^{(\ast)} e^+ e^-)},
\end{align}
are predicted to be $R_{K^{(\ast)}}^{\rm{SM}} = 1.00 \pm 0.01$ in the region $1.1 \le q^2 \le 6\,\mathrm{GeV}^2$, with $q^2$ the dilepton invariant mass squared, within the SM~\cite{Hiller:2003js,Bobeth:2007dw,Bordone:2016gaq,Isidori:2020acz}, while the LHCb measurements both in 2017~\cite{LHCb:2017avl} and 2019~\cite{LHCb:2019hip} exhibited a deviation at $\sim2.5\sigma$ level. Strikingly, the latest update from the LHCb measurement~\cite{LHCb:2021trn}, with
\begin{align}\label{RKnew}
	R_K(1.1 \le q^2 \le 6\,\mathrm{GeV}^2) = 0.846^{+0.042+0.013}_{-0.039-0.012},
\end{align}
has pushed the deviation to be even at the level of $3.1\sigma$, due to the reduced experimental uncertainties. This implies, therefore, a stronger hint of new physics (NP) beyond the SM that violates the lepton-flavor universality (LFU).

The $R_{K^{(\ast)}}$ anomalies have triggered quite a lot of NP proposals under extensive and intensive investigations (see, e.g., the recent reviews~\cite{Albrecht:2021tul,London:2021lfn} and references therein). In particular, the two-Higgs-doublet model (2HDM) extended with right-handed neutrinos~\cite{Iguro:2018qzf,Li:2018rax,Crivellin:2019dun,DelleRose:2019ukt} is an interesting NP candidate, since it can connect the intriguing LFU violation with the neutrino masses---another big mystery in contemporary particle physics domain. Thus far, either Dirac~\cite{Iguro:2018qzf} or Majorana~\cite{Li:2018rax,Crivellin:2019dun,DelleRose:2019ukt} neutrinos have been considered to address the $R_{K^{(\ast)}}$ anomalies. In these scenarios, the dominant NP contribution to the $R_{K^{(\ast)}}$ anomalies is assisted by the right-handed neutrinos running in the box diagrams that are most relevant for the $b\to s \mu^+\mu^-$ transition, and the resulting LFU-violating Wilson coefficients are predicted in the direction, $C_{9\mu}^{\mathrm{NP}}=-C_{10\mu}^{\mathrm{NP}}$, which implies a left-handed NP effect in the muon sector and is persistently favored by the updated global fits following Eq.~\eqref{RKnew}~\cite{Alguero:2019ptt,Alok:2019ufo,Carvunis:2021jga,Angelescu:2021lln,Geng:2021nhg,Cornella:2021sby,Kriewald:2021hfc,Alguero:2021anc,Hurth:2021nsi,Li:2021toq,Altmannshofer:2021qrr,Alok:2022pjb,Alguero:2022wkd}. However, as pointed out in Ref.~\cite{Crivellin:2019dun}, the LFU-conserving Wilson coefficient $C_{10\ell,Z}^\mathrm{NP}$ resulting from the $Z$-penguin diagrams could contribute as large as the LFU-violating ones. Even though the LFU-conserving contribution $C_{10\ell,Z}^\mathrm{NP}$ cannot explain the $R_{K^{(\ast)}}$ anomalies alone, its comparable contribution can affect how the $R_{K^{(\ast)}}$ anomalies are numerically addressed in the direction $C_{9\mu}^{\mathrm{NP}}=-C_{10\mu}^{\mathrm{NP}}$. Therefore, theoretical NP models with comparable $C_{9\mu}^{\mathrm{NP}}=-C_{10\mu}^{\mathrm{NP}}$ and $C_{10\ell,Z}^\mathrm{NP}$ should match a two-parameter global fit following Eq.~\eqref{RKnew}~\cite{Alguero:2021anc}, which have been unfortunately neglected in Refs.~\cite{Iguro:2018qzf,Li:2018rax,Marzo:2019ldg}.

Many previous studies focused on heavy Majorana neutrinos for the $R_{K^{(\ast)}}$ resolution. Nevertheless, as found in Ref.~\cite{Li:2018rax}, the solution of the $R_{K^{(\ast)}}$ anomalies in the direction $C_{9\mu}^{\mathrm{NP}}=-C_{10\mu}^{\mathrm{NP}}$ is insensitive to the Majorana neutrino masses below the electroweak scale. This implies that the difference between heavy and light Majorana neutrinos cannot be simply distinguished by the $R_{K^{(\ast)}}$ resolution. Besides focusing on the Majorana neutrinos, the topics with Dirac neutrinos have also received increasing attention in these years, especially in connection with the phenomenologies~\cite{Steigman:2012ve,Abazajian:2019oqj,Luo:2020sho,Adshead:2020ekg,Luo:2020fdt,Li:2022yna} of the big-bang nucleosynthesis (BBN) and the cosmic microwave background (CMB), as well as the baryon asymmetry of the Universe~\cite{Li:2020ner,Li:2021tlv}. The Dirac neutrino effects in the $b\to s \mu^+\mu^-$ process have been noticed in Ref.~\cite{Iguro:2018qzf}. However, it was found that since $\mathcal{O}(1)$ Dirac neutrino Yukawa couplings are generically required to explain the $R_{K^{(\ast)}}$ anomalies, the thermalized right-handed Dirac neutrinos with such large couplings in the early Universe would make an undesired shift of the effective neutrino number, $\Delta N_{\rm eff}=N_{\rm  eff}-N_{\rm eff}^{\rm SM}$, at the BBN and CMB epochs, where the SM prediction reads $N_{\rm eff}^{\rm SM} = 3.044-3.045$~\cite{Mangano:2005cc,deSalas:2016ztq,Gariazzo:2019gyi,EscuderoAbenza:2020cmq,Akita:2020szl,Froustey:2020mcq,Bennett:2020zkv}. Nevertheless, the above conclusion depends crucially on how many right-handed Dirac neutrinos are thermalized in the early Universe and on the decoupling temperature of thermalized neutrinos, which can result in different levels of the $\Delta N_{\rm eff}$ shift.  
 
In addition to the extensive investigations of the heavy nature of Majorana neutrinos resorted to address the $R_{K^{(\ast)}}$ anomalies, it is also interesting to consider the situations where the eV-scale Majorana neutrinos are included. In this paper, we will consider the NP effects on the $R_{K^{(\ast)}}$ anomalies from either an eV-scale right-handed Majorana or a right-handed Dirac neutrino. While the NP effects arising from these two cases are indistinguishable in terms of the $R_{K^{(\ast)}}$ resolution alone, their impacts on the early Universe are different in generating an observable $\Delta N_{\rm eff}$ shift due to the spinor nature of the neutrinos involved, i.e., the Majorana spinor for the former and the Weyl spinor for the latter, respectively. Therefore, it becomes possible to distinguish these two solutions via the observation of different $\Delta N_{\rm eff}$ shifts in the cosmic regime.

Noticeably, the extra radiation that generates a significant $\Delta N_{\rm eff}$ shift is one of the simplest candidates to mitigate the Hubble ($H_0$) tension (see, e.g., Refs.~\cite{DiValentino:2021izs,Perivolaropoulos:2021jda,Schoneberg:2021qvd,Shah:2021onj,Abdalla:2022yfr} for the latest reviews), which signifies a notorious discrepancy between the local measurements of the present-day Hubble parameter from the SH0ES collaboration~\cite{Riess:2018uxu,Riess:2019cxk,Riess:2020fzl} (according to the publication years, the three references will be dubbed R18, R19, and R20, respectively) and the Planck CMB inferred value under the standard $\Lambda$CDM baseline~\cite{Planck:2018vyg},
\begin{align}
	H_0=(67.4\pm 0.5)~\text{km}\cdot \text{s}^{-1}\cdot \text{Mpc}^{-1}.
\end{align}
The $H_0$ tension is further worsened by the updated SH0ES measurements (R21)~\cite{Riess:2021jrx}, with
\begin{align}
H_0=(73.04\pm 1.04)~\text{km}\cdot \text{s}^{-1}\cdot \text{Mpc}^{-1},
\end{align} 
enhancing the deviation from the Planck 2018 data to $5\sigma$. A plethora of investigations have invoked a shift of the effective neutrino number, $\Delta N_{\rm eff}\simeq 1.0$, to \textit{address} the $H_0$ tension~\cite{Carneiro:2018xwq,Kreisch:2019yzn,Vagnozzi:2019ezj,Franchino-Vinas:2021nsf}. As illustrated in Ref.~\cite{Vagnozzi:2019ezj}, an extra free $N_{\rm eff}$ beyond the original six $\Lambda$CDM parameters can make a genuine shift in the central value of $H_0$ from Planck measurements, and the $H_0$ tension can be relieved with $N_{\rm eff} \approx 3.95$. Here, $N_{\rm eff}$ serves as a NP source to shift the $\Lambda$CDM predictions inferred from CMB, BAO and Pantheon Supernovae Type-Ia data to be in agreement with the local $H_0$ measurements. Being different from estimating $N_{\rm eff}$ simply by combining the high-redshift measurements with the local $H_0$ data in many other works, the data-analyzing method proposed in Ref.~\cite{Vagnozzi:2019ezj} opens a new avenue to ease the $H_0$ tension. However, such a too large shift is disfavored by the high-$\ell$ Planck CMB polarization measurements~\cite{Bernal:2016gxb,Planck:2018vyg,RoyChoudhury:2020dmd,Aloni:2021eaq}. More recent analyses show, instead, that a shift of $0.2<\Delta N_{\rm eff}<0.6$ is able to \textit{ease} the $H_0$ tension. For instance, Ref.~\cite{Seto:2021xua} points out two possible regimes with/without BBN data,
\begin{align}\label{R19noBBN}
	&3.22<N_{\rm eff}<3.49~(68\%\,\text{CL})\qquad  \text{for CMB+BAO+Pantheon+R19},
  \\[0.2cm]
	&3.16<N_{\rm eff}<3.40~(68\%\,\text{CL})\qquad \text{for CMB+BAO+Pantheon+R19+BBN},\label{R19BBN}
\end{align}
in which the SH0ES 2019 measurements (R19)~\cite{Riess:2019cxk} are included. These patterns are also consistent with that observed in Ref.~\cite{Seto:2021tad} where an additional electron-type lepton asymmetry $\xi_e$ in the neutrino sector is introduced, giving 
\begin{equation}
 N_{\rm eff}=3.46\pm 0.13~(68\%\,\text{CL}), \quad \xi_e=0.04  \quad \text{for CMB+BAO+Pantheon+R19+BBN},
\end{equation}
with a larger central value of $N_{\rm eff}$ than from Eq.~\eqref{R19BBN}. Intriguingly, the introduction of a lepton asymmetry is also supported by the very recently probed anomaly in the helium-4 abundance~\cite{Matsumoto:2022tlr}, which results in~\cite{Matsumoto:2022tlr,Kawasaki:2022hvx,Burns:2022hkq}
\begin{align}\label{newBBN}
	N_{\rm eff}=3.22^{+0.33}_{-0.30}, \qquad \xi_e=0.05\pm0.03.
\end{align}
It is noted that all the central values of $N_{\rm eff}$ obtained above are larger than the previous CMB+BBN result~\cite{Fields:2019pfx}, $N_{\rm eff}=2.843\pm 0.154$.
Therefore, it can be inferred from Eqs.~\eqref{R19noBBN}--\eqref{newBBN} that an increased $N_{\rm eff}$ will be helpful to mitigate the $H_0$ tension, though an updated analysis of the combined data set from CMB+BAO+Pantheon+R21+BBN is currently not available. Other possible patterns, such as the extra radiation in the presence of additional non-free-streaming degrees of freedom (d.o.f)~\cite{Brust:2017nmv,Blinov:2020hmc}, also found that comparable $N_{\rm eff}$ values are favored to ease the $H_0$ tension. Recently, the mitigation of the $H_0$ tension with $\Delta N_{\rm  eff}\simeq \mathcal{O}(0.5)$ has been studied in some explicit models~\cite{Escudero:2019gzq,Escudero:2019gvw,Aloni:2021eaq,Aboubrahim:2022gjb}.

The lesson learned from above suggests that a full resolution of the $H_0$ tension could be a result of multidisciplinary interplay, in which the extra radiation serves as a fractional but important role. In relating the observed anomalies in particle physics domain, it is compelling to consider the situation where the underlying mechanism for the $\Delta N_{\rm eff}$ shift is \textit{naturally} provided by the $R_{K^{(\ast)}}$ resolution via an eV-scale right-handed Majorana or a right-handed Dirac neutrino, which motivates our present study. In this paper, we will show that such a connection can indeed be realized in a flavor-specific 2HDM framework, where only one right-handed Majorana or Dirac neutrino has significant interactions with the extra Higgs bosons present in the model.  

The paper is organized as follows. We begin in Sec.~\ref{model} with a description of the framework, dubbed $t\nu$2HDM, and then take into account in Sec.~\ref{pheno} the most relevant constraints from the low-energy flavor physics, the perturbative unitarity condition, as well as the LHC direct searches. In Sec.~\ref{contri} we will discuss the NP contributions to the $R_{K^{(\ast)}}$ anomalies and the mitigation of the $H_0$ tension. Then, we present in Sec.~\ref{num} our detailed numerical analyses of the viable parameter space for the $R_{K^{(\ast)}}$ resolution, as well as the correlation between the $R_{K^{(\ast)}}$ anomalies and the $H_0$ tension. Conclusions are finally made in Sec.~\ref{con}.

\section{Flavor-specific two-Higgs-doublet model}
\label{model}
The 2HDM is a simple extension of the SM by adding a second Higgs doublet to the SM particle content~\cite{Lee:1973iz,Branco:2011iw}. Any specific 2HDM framework is characterized by its Yukawa interactions and scalar potential, both of which can be either specified by some symmetry backgrounds or by purely phenomenological considerations. For our purpose to address the $R_{K^{(\ast)}}$ anomalies with a link to the $H_0$ tension, here we follow a data-driven approach.

\subsection{Quasi-degenerate Higgs mass spectrum}
\label{subsec:massspectrum}
The two Higgs doublets $H_{1,2}$ in the model are constructed in the so-called Higgs basis~\cite{Lavoura:1994fv,Branco:2011iw} as
\begin{equation}\label{higgs}
	H_1 =
	\begin{pmatrix}
		G^+
		\\
		(v + \phi_1 + i G^0)/\sqrt{2} 
	\end{pmatrix}
	, \qquad H_2 =
	\begin{pmatrix}
		H^+
		\\
		(\phi_2 + i A)/\sqrt{2} 
	\end{pmatrix},
\end{equation}
where the vacuum expectation value $v\simeq 246\,\mathrm{GeV}$ is responsible for generating the fermion and gauge-boson masses, and $G^{+,0}$ are the Goldstone bosons. Here we will assume a CP-conserving Higgs potential~\cite{Branco:2011iw}. Then, $H^+$ and $A$ are the physical charged and neutral pseudo-scalar Higgs bosons respectively, while the neutral scalars $\phi_{1,2}$ are the superposition of the two mass eigenstates $H^0$ and $h$, which can be, in general, written as 
\begin{align}
  \phi_1=h\cos\theta +H^0\sin\theta , \qquad \phi_2=-h\sin\theta +H^0\cos\theta ,
\end{align}  
with the mixing angle $\theta$ determined completely by the parameters in the Higgs potential. Given that $\cos\theta\approx 1$ is favored by the current LHC data on various SM-like Higgs signals (see, e.g., Refs.~\cite{Chowdhury:2017aav,Haller:2018nnx} for the recent global fits of the 2HDMs), which corresponds to the so-called alignment limit, we will consider the case where $H_1$ is the SM Higgs doublet such that $h$ corresponds to the observed Higgs boson~\cite{ATLAS:2012yve,CMS:2012qbp}, while $H_2$ is the NP doublet with $H^0$ corresponding to the extra physical neutral scalar. 

In the alignment limit, the Higgs potential $V(H_1, H_2)$ can be readily constructed in terms of the free potential parameters governing the Higgs mass spectrum. In principle, these free parameters receive various theoretical and phenomenological constraints, such as the vacuum stability, perturbative unitarity, electroweak precision tests, as well as collider direct detection~\cite{Chowdhury:2017aav,Haller:2018nnx}. Nevertheless, the mass spectrum of the physical states $H^+$, $H^0$ and $A$ is still undetermined by the current LHC direct searches. In particular, a quasi-degenerate Higgs mass spectrum,
\begin{equation}\label{eq.Hmass}
	m_S \equiv m_{H^0} \simeq m_A \simeq m_{H^+},
\end{equation}
still remains as a possible regime, and will be concerned here.

\subsection{Flavor-specific Yukawa structure}
\label{subsec:yukawastructure}
In order to address the $R_{K^{(\ast)}}$ anomalies with a link to the $H_0$ tension, we will consider the following Yukawa interactions:
\begin{align}\label{Lag}
	\mathcal{L}_Y &= \mathcal{L}_Y^{\rm{SM}}(H_1) +\mathcal{L}_Y (H_2),
	\\[0.2cm]
	\mathcal{L}_Y (H_2) &=- X_u \bar{Q}_L \tilde{H}_2 u_R - X_\nu \bar{E}_L \tilde{H}_2 \nu_R + \text{h.c.},\label{Lag2}
\end{align}
where $\mathcal{L}_Y^{\rm{SM}}(H_1)$ denotes the SM Yukawa Lagrangian associated with the Higgs doublet $H_1$, while $\mathcal{L}_Y (H_2)$ encodes the NP interactions related to the second Higgs doublet $H_2$, with $\tilde{H}_2=i\sigma_2 H_2^\ast$ and $\sigma_2$ being the second Pauli matrix. The left-handed fermion doublets $\bar Q_L$ and $\bar E_L$ are specified, respectively, as
\begin{equation}
	\bar Q_L\equiv (\bar u_L, \bar d_L V^\dagger), \qquad \bar E_L\equiv (\bar \nu_L U^\dagger, \bar e_L ),
\end{equation}	
where all the chiral fermions $f_{L,R}$~($f=u,d,e,\nu$) are the physical fields, with $V$ and $U$ representing the Cabibbo–Kobayashi–Maskawa (CKM) and Pontecorvo–Maki–Nakagawa–Sakata (PMNS) matrices, respectively. For the Yukawa matrices $X_{u,\nu}$, we propose the following phenomenologically viable structure: 
\begin{align}\label{Yukawa}
	X_{u,ij}=\kappa_t \delta_{i3}\delta_{j3}, \qquad X_{\nu,ij}=\kappa_\nu \delta_{i2}\delta_{js},
\end{align} 
where $\kappa_{t,\nu}$ are the only nonzero \textit{real} effective couplings, and the flavor index $s$ characterizes the one-flavor right-handed neutrino that couples to the muon lepton in the charged scalar current. It should be noted that the explicit right-handed neutrino flavor is irrelevant here and will be simply denoted as $\tilde{\nu}_{R}$ hereafter.
	
Our proposal of Eq.~\eqref{Yukawa} comes from various data-driven considerations. In the quark sector, Eq.~\eqref{Lag2} together with Eq.~\eqref{Yukawa} would induce only neutral scalar currents associated with the top quark, and the charged scalar interactions,
\begin{align}
	\bar d_{L,i} V^*_{ki} X_{u,kj} u_{R,j} H^-+\text{h.c.}, 
\end{align} 
have only significant effects on the third generation of quarks due to the hierarchical structure of the CKM matrix. These patterns comply with the current observation that only significant NP contributions are allowed in the third generation and the flavor-changing neutral scalar currents are severely constrained by the experimental data~\cite{Crivellin:2013wna,Glashow:2014iga,Calibbi:2015kma}. In the lepton sector, on the other hand, Eqs.~\eqref{Lag2} and \eqref{Yukawa} would indicate that there are only neutral scalar currents in the neutrino sector, while the charged scalar interaction is only stimulated by the one-flavor right-handed neutrino $\tilde{\nu}_{R}$ that has a significant coupling to the muon lepton, namely
\begin{align}\label{muonphilic}
\kappa_{\nu} \bar \mu_L \tilde \nu_R H^- + \text{h.c.}
\end{align}  
Such particular patterns follow closely the tight bounds from the charged lepton-flavor violating processes $\ell_i\to \ell_j \gamma$ mediated by the right-handed neutrino at the loop level~\cite{Li:2018rax} and the muon decay $\mu\to e \nu \bar{\nu}$ mediated by the charged Higgs at the tree level. Furthermore, the reason for allowing only one rather than two or three flavors of right-handed neutrinos to interact with the muon lepton comes from the observation that, if more than one right-handed neutrinos have significant couplings to account for the $R_{K^{(\ast)}}$ anomalies, the resulting parameter space will readily force them to establish thermal equilibrium in the early Universe and thus generate an unacceptably large $\Delta N_{\rm eff}$ shift~\cite{Iguro:2018qzf,Abazajian:2019oqj}, as will also be confirmed in this paper. Finally, it should be emphasized that, since we are here interested in the connection between the $R_{K^{(\ast)}}$ anomalies and the $H_0$ tension via a minimal setup, other couplings not concerned in $X_{u,\nu}$ are not necessary to vanish strictly, but rather signify the meaning of \textit{phenomenological smallness} in their own right. Besides, we will not concern here the symmetry underlying such a flavor-specific Yukawa structure given by Eq.~\eqref{Yukawa}, though interesting possibilities, such as the Branco-Grimus-Lavoura-based scenarios~\cite{Branco:1996bq} and the mass-powered-like textures~\cite{Li:2019xmi}, may deserve further exploitation.
	
The above considerations result in our flavor-specific 2HDM framework that points toward significant NP effects associated with the top quark $t$ and the one-flavor right-handed neutrino $\tilde{\nu}_R$, and will be therefore dubbed the $t\nu$2HDM hereafter. As mentioned in Sec.~\ref{intro}, the neutrino nature, being of Majorana or of Dirac type, is unspecified by  the $R_{K^{(\ast)}}$ resolution alone. However, when $\tilde{\nu}_R$ is an eV-scale Majorana neutrino, its impact on the $\Delta N_{\rm eff}$ shift will be different from that with a Dirac neutrino, especially when the shift is linked to the mitigation of the $H_0$ tension. Furthermore, if $\tilde{\nu}_R$ is of Majorana type, it can be embedded into the seesaw mechanism (see, e.g., Refs.~\cite{Mohapatra:2005wg,Xing:2020ijf} for the recent comprehensive reviews) where two more right-handed Majorana neutrinos are introduced with the presence of a Majorana mass term,  
\begin{align}\label{Majmass}
	\frac{1}{2} \overline{(\nu_R)^c} M_R \nu_R + \text{h.c.},
\end{align}
where $(\nu_R)^c=C\overline{\nu_R}^T$, and $C=i\gamma^2\gamma^0$ is the charge conjugation matrix. Then, the active neutrino masses are generated via the seesaw formula,
\begin{align}\label{ssform}
 m_{\nu,ij}\simeq -\frac{v^2}{M_{R,k}} Y_{\nu,ik}Y_{\nu,jk},
\end{align}
in the basis where the symmetric matrix $M_R$ is already diagonal. Here $Y_\nu$ is the neutrino Yukawa matrix from $\mathcal{L}_Y^{\rm{SM}}(H_1)$ in Eq.~\eqref{Lag}, i.e.,
\begin{align}\label{nuYukawa}
	Y_\nu \bar{E}_L \tilde{H}_1 \nu_R +\text{h.c.},
\end{align}
encoding the active-sterile neutrino mass mixing and the mixing-induced interactions via~\cite{Ibarra:2010xw} 
\begin{align}
W_{\nu_{},ij}^*\simeq\frac{v}{M_{R,j}} Y_{\nu,ij}.
\end{align}
An important observation arises if one of the sterile neutrino eigenstates is at the eV-scale, say $M_{R,1}\simeq \mathcal{O}(\mathrm{eV})$: To avoid the constraints on the active-sterile neutrino mixing~\cite{Fernandez-Martinez:2016lgt} and in particular for an eV-scale sterile neutrino~\cite{Dasgupta:2021ies}, the first column of $Y_{\nu}$ must be strongly suppressed,  because otherwise $W_{\nu,i1}$ would be enhanced by a factor of $v/M_{R,1}\simeq 10^{11}$. In the asymptotically safe limit, $Y_{\nu,i1}=0$, the active neutrino mass matrix from Eq.~\eqref{ssform} would be of rank two, making the lightest active neutrino massless. Therefore, if the eV-scale $\tilde{\nu}_R$ belongs to the lightest sterile neutrino in the seesaw mechanism, the lightest active neutrino in the $3\nu$ oscillation paradigm~\cite{Capozzi:2021fjo} is essentially massless. Further constraints on $W_{\nu,ij}$ will not be discussed in the following as the NP effects concerned in this paper do not rely on the neutrino Yukawa matrix $Y_{\nu}$. If $\tilde \nu_R$ is of Dirac type, on the other hand, the Dirac neutrino mass may also be generated via Eq.~\eqref{nuYukawa} but with the absence of the Majorana mass term given by Eq.~\eqref{Majmass}. In either case, Eq.~\eqref{Lag2} will encode all the NP interactions concerned in this paper.  

In the following sections, we will consider important constraints from flavor and collider physics, as well as the perturtive unitarity on the $t\nu$2HDM framework, which is only  characterized by the three free parameters,
\begin{align}
\kappa_t,\quad \kappa_\nu, \quad m_S,
\end{align} 
and show the viable parameter space in addressing the $R_{K^{(\ast)}}$ anomalies. We will further show that the resulting favored parameter space induces a shift $\Delta N_{\rm eff}\simeq 1.0$ in the Dirac neutrino case and $\Delta N_{\rm eff}\simeq 0.5$ in the Majorana case, respectively.

\section{Phenomenological and theoretical constraints}
\label{pheno}
The $t\nu$2HDM framework signifies significant NP effects associated with the third generation of quarks and the muon lepton. In this section, we will discuss the most relevant constraints on the model from the low-energy flavor physics, the perturbative unitarity condition, and the LHC direct searches. 

\subsection{Constraints from the low-energy flavor physics}

\subsubsection{$\bar{B}\to X_s\gamma$}
In the framework of low-energy effective field theory, the effective Hamiltonian governing the radiative $b\to s\gamma$ decay at the scale $\mu_b \simeq \mathcal{O}(m_b)$ reads 
\begin{equation}
\mathcal{H}_{\rm eff}(b\to s\gamma) = -\frac{4G_F}{\sqrt{2}} V_{ts}^\ast V_{tb} \left[ \sum_{i=1}^6 C_i(\mu_b) \mathcal{O}_i + C_{7\gamma}^{(\prime)}(\mu_b) \mathcal{O}_{7\gamma}^{(\prime)} + C_{8g}^{(\prime)}(\mu_b) \mathcal{O}_{8g}^{(\prime)} \right],
\end{equation}
where $G_F=1/(\sqrt{2}v^2)$ is the Fermi constant, and the terms proportional to $V_{us}^\ast V_{ub}$ have been neglected in view of $|V_{us}^\ast V_{ub}/V_{ts}^\ast V_{tb}|<0.02$. The explicit expressions of the current-current ($\mathcal{O}_{1,2}$) and QCD-penguin ($\mathcal{O}_{3-6}$) operators can be found, e.g., in Refs.~\cite{Grinstein:1990tj,Borzumati:1998tg,Bobeth:1999ww,Degrassi:2010ne}, while the magnetic dipole operators are defined, respectively, by
\begin{equation}
\mathcal{O}_{7\gamma}^{(\prime)} \equiv \frac{e}{16\pi^2} m_b (\bar{s} \sigma^{\mu \nu} P_{R(L)} b) F_{\mu \nu} , \qquad \mathcal{O}_{8g}^{(\prime)} \equiv \frac{g_s}{16\pi^2} m_b (\bar{s} \sigma^{\mu \nu} T^a P_{R(L)} b) G^a_{\mu \nu},
\end{equation}
where $P_{R,L}=(1\pm\gamma_5)/2$ are the right- and left-handed chirality projectors.

\begin{figure}[t]
	\centering
	\includegraphics[scale=0.6]{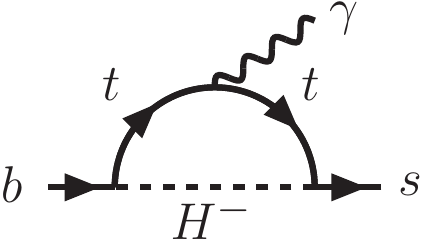} \qquad
	\includegraphics[scale=0.6]{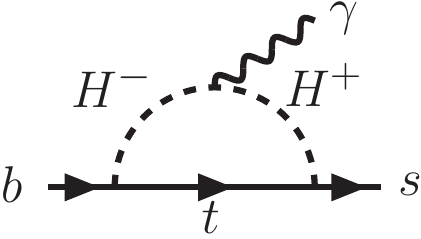} \qquad
	\includegraphics[scale=0.6]{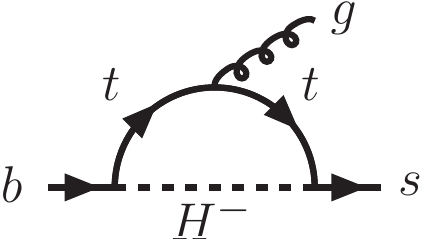} 
	\caption{The relevant NP photon- (the first two) and gluon-penguin (the last) diagrams contributing to the inclusive radiative $\bar{B}\to X_s\gamma$ decay.}\label{fig.bsg}
\end{figure}

In the $t\nu$2HDM framework, the NP contributions to the Wilson coefficients $C_{1-6}$ are absent, and their contributions to the primed dipole coefficients $C_{7\gamma,8g}^\prime$ are suppressed by the ratio $m_s/m_b$. As a consequence, the dominant NP influence on the $b\to s\gamma$ transition stems from the unprimed $C_{7\gamma}$ and $C_{8g}$. After a direct evaluation of the one-loop penguin diagrams with the charged Higgs running in the loop, as shown in Fig.~\ref{fig.bsg}, we can obtain the NP Wilson coefficients at the matching scale $\mu_S\simeq \mathcal{O}(m_S)$~\cite{Borzumati:1998tg,Bobeth:1999ww,Degrassi:2010ne}:
\begin{equation}\label{C7C8muS}
C_{7\gamma}^{\rm{NP}} (\mu_S) = \frac{\sqrt{2} \kappa_t^2}{4G_F m_S^2} E_{7\gamma}^{\rm{NP}}, \qquad 
C_{8g}^{\rm{NP}} (\mu_S) = \frac{\sqrt{2} \kappa_t^2}{4G_F m_S^2} E_{8g}^{\rm{NP}},
\end{equation} 
where the scalar functions are defined, respectively, by 
\begin{align}
E_{7\gamma}^{\rm{NP}} &= \frac{-7+12z_t+3z_t^2-8z_t^3+6z_t(3z_t-2)\ln z_t}{72(1-z_t)^4},\\
E_{8g}^{\rm{NP}} &= \frac{-2-3z_t+6z_t^2-z_t^3-6z_t\ln z_t}{24(1-z_t)^4},
\end{align}
with $z_t \equiv m_t^2/m_S^2$ and $m_t$ the top-quark $\overline{\rm{MS}}$ mass.

To evaluate the NP contributions to the branching ratio $\mathcal{B}(\bar{B}\to X_s\gamma)$, we have to run the Wilson coefficients from the matching scale $\mu_S$ down to the low-energy scale $\mu_b$~\cite{Buras:2011zb,Blanke:2011ry}. Generically, the Wilson coefficient $C_{7\gamma}(\mu_b)$ can be divided into two parts,
\begin{equation}\label{eq.c7}
	C_{7\gamma}(\mu_b) = C_{7\gamma}^{\rm{SM}}(\mu_b) + C_{7\gamma}^{\rm{NP}}(\mu_b),
\end{equation}
which contributes to $\mathcal{B}(\bar{B}\to X_s\gamma)$ with a photon-energy cutoff $E_\gamma < 1.6\,\mathrm{GeV}$ via~\cite{Buras:2011zb,Blanke:2011ry} 
\begin{equation}\label{BXs.the}
	\mathcal{B}(\bar{B}\to X_s\gamma)_{E_\gamma <1.6\,\mathrm{GeV}} = R ~|C_{7\gamma}(\mu_b)|^2.
\end{equation}
Here the overall factor reads numerically $R = 2.47 \times 10^{-3}$, and we have neglected the small non-perturbative effect. The SM contribution $C_{7\gamma}^{\rm{SM}}(\mu_b)$ has been calculated up to the next-to-next-to-leading order in QCD~\cite{Misiak:2004ew,Gorbahn:2005sa,Czakon:2006ss}, and the resulting numerical value reads~\cite{Misiak:2015xwa,Misiak:2020vlo}
\begin{equation}
C_{7\gamma}^{\rm{SM}}(\mu_b) = -0.371 \pm 0.009,
\end{equation}
while the NP part $C_{7\gamma}^{\rm{NP}}(\mu_b)$ is given by  
\begin{equation}\label{eq.c7np}
C_{7\gamma}^{\rm{NP}}(\mu_b) = \kappa_7 ~C_{7\gamma}^{\rm{NP}}(\mu_S) + \kappa_8 ~C_{8g}^{\rm{NP}} (\mu_S),
\end{equation}
where $C_{7\gamma}^{\rm{NP}}(\mu_S)$, $C_{8g}^{\rm{NP}} (\mu_S)$ are already given by Eq.~\eqref{C7C8muS}, and the magic numbers are evaluated to be $\kappa_7 =0.457$, $\kappa_8 = 0.125$ at the NP scale $\mu_S \sim \mathcal{O}(1)\,\mathrm{TeV}$.

By comparing the theoretical prediction given by Eq.~\eqref{BXs.the} with the current world-averaged experimental data~\cite{HFLAV:2019otj},
\begin{equation}\label{BXs.exp}
\mathcal{B}(\bar{B}\to X_s\gamma)_{E_\gamma <1.6\,\mathrm{GeV}}^{\rm{exp}} = (3.32 \pm 0.15) \times 10^{-4},
\end{equation}
we can set bounds on the NP Wilson coefficients presented in Eq.~\eqref{C7C8muS} and the allowed parameter space $(\kappa_t,m_S)$ can be, therefore, extracted. In Sec.~\ref{num}, we will apply the $\mathcal{B}(\bar{B}\to X_s\gamma)$ constraint within the $1\sigma$ uncertainty.

\subsubsection{$B^0_{d,s}-\bar{B}^0_{d,s}$ mixings}
\label{BBmixing}
Next, we turn our attention to the mass differences $\Delta M_{d,s}$ describing the strength of the $B^0_{d,s}-\bar{B}^0_{d,s}$ mixings. The theoretical description of the $B^0_{d,s}-\bar{B}^0_{d,s}$ mixings can be realized in terms of the low-energy effective Hamiltonian
\begin{equation}
\mathcal{H}_{\rm eff}^{\Delta B=2} = \frac{G_F^2}{16\pi^2} m_W^2 (V_{tb}^\ast V_{tq})^2 \left[ \sum_{i=1}^5 C_{iq}(\mu_b) \mathcal{Q}_{iq} + \sum_{i=1}^3 \tilde{C}_{iq}(\mu_b) \tilde{\mathcal{Q}}_{iq} \right] + \text{h.c.},
\end{equation}
where $m_W$ is the $W$-boson mass, and $q=d(s)$ for the neutral $B_{d(s)}$ meson. The explicit expressions of the four-quark operators can be found, e.g., in Refs.~\cite{Buras:2001ra,Becirevic:2001jj}.

\begin{figure}[t]
	\centering
	\includegraphics[scale=0.6]{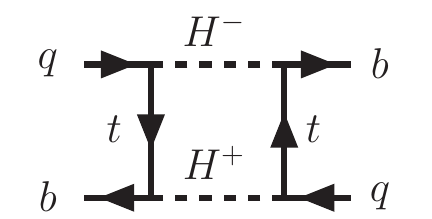}\qquad
	\includegraphics[scale=0.6]{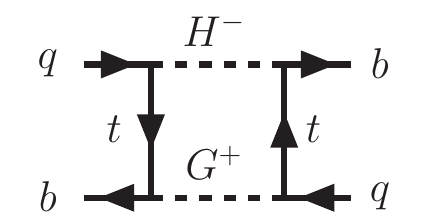}\qquad
	\includegraphics[scale=0.6]{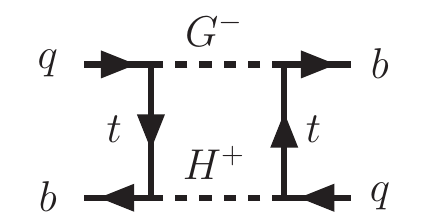} \\[0.35cm]
	\includegraphics[scale=0.6]{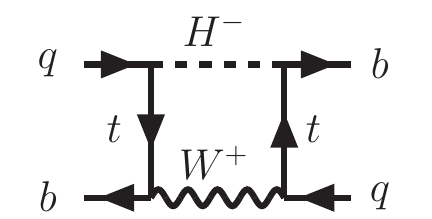}\qquad
	\includegraphics[scale=0.6]{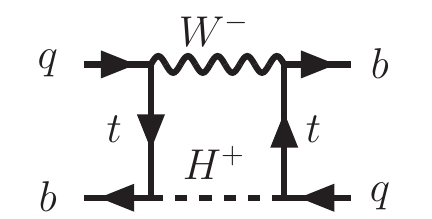}
	\caption{The relevant NP box diagrams responsible for the $B^0_{q}-\bar{B}^0_{q}$ mixing in the 't Hooft-Feynman gauge.} \label{fig.bbar}
\end{figure}

In both the SM and the $t\nu$2HDM framework, the only significant Wilson coefficient responsible for the neutral $B$-meson mixing comes from $C_{1q} (\mu_b)$, which corresponds to the four-quark operator
\begin{equation}
\mathcal{Q}_{1q}= (\bar{b}_\alpha \gamma_\mu P_L q_\alpha) (\bar{b}_\beta \gamma^\mu P_L q_\beta),
\end{equation}
where the Greek letters $\alpha$ and $\beta$ denote the quark color indices. The mass difference of the neutral $B$-meson mixing can be expressed in terms of the off-diagonal matrix element, $\Delta M_q = 2|\mathcal{M}_{12}^q|$, with the latter given by~\cite{Buras:2001ra,Becirevic:2001jj}
\begin{equation}
\left[\mathcal{M}_{12}^q\right]^\ast = \langle \bar{B}_q^0| \mathcal{H}_{\rm eff}^{\Delta B=2} |B_q^0 \rangle = \frac{G_F^2}{16\pi^2} m_W^2 (V_{tb}^\ast V_{tq})^2 C_{1q} (\mu_b) \langle \bar{B}_q^0| \mathcal{Q}_{1q} |B_q^0 \rangle.
\end{equation}
Here the hadronic matrix element $\langle \bar{B}_q^0| \mathcal{Q}_{1q} |B_q^0 \rangle$ encodes the non-perturbative QCD effect, while the perturbative contribution is absorbed into the short-distance Wilson coefficient $C_{1q} (\mu_b)$. Normalizing the NP to the SM contribution, we can parameterize the theoretical prediction of $\Delta M_q$ as
\begin{equation}
\Delta M_q = \Delta M_q^{\rm{SM}} \left|1+ \Delta^{\rm{NP}}_q \right|,
\end{equation}
where $\Delta M_q^{\rm{SM}}$ denotes the SM contribution. For a NP scale of $\mathcal{O}(1)\,\mathrm{TeV}$ concerned in this paper, the correction $\Delta^{\rm{NP}}_q$ is given by
\begin{equation}
\Delta^{\rm{NP}}_q = U^{(0)}(\mu_W,\mu_t) U^{(0)}(\mu_t,\mu_S) \frac{C_{1q}^{\rm{NP}}(\mu_S)}{C_{1q}^{\rm{SM}}(\mu_W)},
\end{equation}
where $U^{(0)}(\mu_i,\mu_j)$ represents the leading-order QCD evolution function from the high-scale $\mu_j$ to the low-scale $\mu_i$~\cite{Buras:2001ra}, and $C_{1q}^{\rm{SM}}(\mu_W)$ is the SM Wilson coefficient evaluated at $\mu_W \simeq \mathcal{O}(m_W)$. Here we have taken into account the threshold effect when evolving across the top-quark mass scale $\mu_t\simeq \mathcal{O}(m_t)$~\cite{Buras:2001ra}. The NP contribution $C_{1q}^{\rm{NP}}(\mu_S)$ is obtained by evaluating the box diagrams shown in Fig.~\ref{fig.bbar}, leading to 
\begin{equation}\label{C1q}
C_{1q}^{\rm{NP}} (\mu_S) = C_{1q}^{H-H}(\mu_S) + C_{1q}^{H-G} (\mu_S) + C_{1q}^{H-W} (\mu_S),
\end{equation}
in which the different parts are given, respectively, as 
\begin{align}
C_{1q}^{H-H} (\mu_S) &\,=\, \frac{\kappa_t^4}{8G_F^2 m_S^4} \mathcal{I}(z_t,z_W),\\
C_{1q}^{H-G} (\mu_S) &\,=\, \frac{\kappa_t^2}{\sqrt{2}G_F m_S^2} \mathcal{J}(z_t,z_W),\\
C_{1q}^{H-W} (\mu_S) &\,=\, \frac{2\sqrt{2}\kappa_t^2 m_W^2}{G_F m_S^4} \mathcal{K}(z_t,z_W),
\end{align}
where $z_W \equiv m_W^2/m_S^2$, and we have introduced the following scalar functions:
\begin{align}
\mathcal{I}(z_t,z_W) &= \frac{1-z_t^2+2z_t\ln z_t}{z_W(1-z_t)^3},\\
\mathcal{J}(z_t,z_W) &= \frac{-z_t^2}{z_W(1-z_t)(z_t-z_W)} + \frac{z_t z_W \ln(z_t/z_W)}{(1-z_W)(z_t-z_W)^2} - \frac{z_t\ln z_t}{z_W(1-z_W)(1-z_t)^2},\\
\mathcal{K}(z_t,z_W) &= \frac{z_t}{z_W(1-z_t)(z_t-z_W)} - \frac{z_t \ln(z_t/z_W)}{(1-z_W)(z_t-z_W)^2} + \frac{z_t\ln z_t}{z_W(1-z_W)(1-z_t)^2}.
\end{align}

The current world-averaged experimental measurements are given, respectively, by~\cite{HFLAV:2022pwe}
\begin{equation}\label{eq:expdeltam}
\Delta M_d^{\mathrm{exp}} = 0.5065 \pm 0.0019~{\mathrm{ps}}^{-1}, \qquad \Delta M_s^{\mathrm{exp}} = 17.765 \pm 0.006~{\rm{ps}}^{-1},
\end{equation}
both of which carry much smaller uncertainties than from the corresponding theoretical predictions~\cite{DiLuzio:2017fdq,Dowdall:2019bea,DiLuzio:2019jyq,Lenz:2019lvd}. In particular, based on the bag parameters calculated in full four-flavor lattice QCD for the first time, the HPQCD collaboration found that~\cite{Dowdall:2019bea}
\begin{equation} \label{eq:HPQCD}
\Delta M_d^{\rm{SM}} = 0.555^{+0.040}_{-0.062}~{\rm{ps}}^{-1} , \qquad \Delta M_s^{\rm{SM}} = 17.59^{+0.85}_{-1.45}~{\rm{ps}}^{-1},
\end{equation}
in which the central value of $\Delta M_d^{\rm{SM}}$ is larger than the experimental data. This in turn implies a discrepancy for the ratio $\Delta M_d/\Delta M_s$ at $\sim 1.7\sigma$. On the other hand, an earlier computation based on the most accurate numerical inputs at that time found that~\cite{DiLuzio:2017fdq}
\begin{align}
\Delta M_s^{\rm{SM}} = (20.01\pm 1.25)~{\rm{ps}}^{-1},
\end{align}
the central value of which is $\sim 1.8\sigma$ above the experimental one given by Eq.~\eqref{eq:expdeltam}. Such a difference has profound implications for NP models that predict sizable positive contributions to $B^0_{s}-\bar{B}^0_{s}$ mixing~\cite{DiLuzio:2017fdq}. While the discrepancies observed in $\Delta M_{d,s}$ are not conclusive yet due to the large theoretical uncertainties, it is interesting to note that an excess over the SM predictions cannot be made reconciled with the $R_{K^{(\ast)}}$ resolution in the $t\nu$2HDM framework, since the NP effects on $\Delta M_{d,s}$ are always positive as can be seen from Eq.~\eqref{C1q}. Therefore, if confirmed with more precise experimental measurements and theoretical predictions, the discrepancy will entail additional NP sources beyond the minimal setup considered in this work. 

In view of the above observations, we will apply in this work the HPQCD results for $\Delta M_{d,s}$ given by Eq.~\eqref{eq:HPQCD} as constraints, but varying the uncertainties within $3\sigma$ conservatively. We would like to emphasize again that the constraining power from $\Delta M_{d,s}$ can be much efficient only when the theoretical uncertainties from, e.g., the $B$-meson decay constants, the bag parameters, and the CKM elements are significantly reduced~\cite{DiLuzio:2017fdq}.

\subsubsection{$K_{S,L}\to\mu^+\mu^-$ decays and $K^0$-$\bar{K}^0$ mixing}
Besides the $B$-meson observables discussed above, the $t\nu$2HDM also makes impacts on the $K$-meson observables, such as the branching ratios of $K_{S,L}\to\mu^+\mu^-$ decays as well as the mass difference $\Delta M_K$ and the $\epsilon_K$ parameter of $K^0$-$\bar{K}^0$ mixing. However, as the kaons are composed of two light quarks, i.e., the up (down) and strange quarks, while the NP interactions in the quark sector within our framework always connect with the top quark~(see Eq.~\eqref{Yukawa}), their leading contributions to the $K$ decays and mixing must stem firstly from the one-loop diagrams with the top quark and the charged Higgs running in the loop. This implies that the NP impacts on the $K$-meson observables are suppressed by both the loop factor and these heavy particle masses, as well as the CKM entries involved. 

Explicitly, we have evaluated the short-distance NP contributions to the branching ratios of $K_{L,S}\to \mu^+\mu^-$ decays, and found that they only result in a negligible effect on the branching ratios of $K_{L,S}\to \mu^+\mu^-$ decays, especially when the sign of the long-distance contribution is chosen to be destructive with the short-distance part~\cite{Isidori:2003ts,DAmbrosio:2017klp,Hou:2022qvx}. We have also checked if the resulting parameter space of the $t\nu$2HDM complies with the constraint from $K^0-\bar{K}^0$ mixing. To this end, fixing the free parameters at a typical benchmark point $(\kappa_t, m_S) \sim (1.0, 1000\,{\rm GeV})$, we find numerically a much weaker impact on the $K^0-\bar{K}^0$ mixing, compared to that obtained through a global fit study~\cite{Crivellin:2013wna}. Thus, these observations promote us to conclude safely that the $K$-meson observables do not put any significant constraints on the $t\nu$2HDM, compared to that obtained from the $B$-meson observables. 

As a consequence, we will not show the constraints from $K$-meson observables in the following numerical analysis.

\subsubsection{LFU tests via $Z$- and $W$-boson decays}
Let us now consider the constraints from the LFU ratios of the di-lepton decays of $Z$ and $W$ gauge bosons, $\Gamma(Z\to \mu^+\mu^-)/\Gamma(Z\to \ell^+\ell^-)$ and $\Gamma(W\to \mu \bar{\nu})/\Gamma(W\to \ell \bar{\nu})$, where $\ell=e$ or $\tau$. For both of these two cases, by encoding the one-loop NP corrections into the renormalized effective vertex in the on-shell scheme, we can readily derive the NP contributions to these LFU ratios. 

\begin{figure}[t]
	\centering
	\includegraphics[scale=0.5]{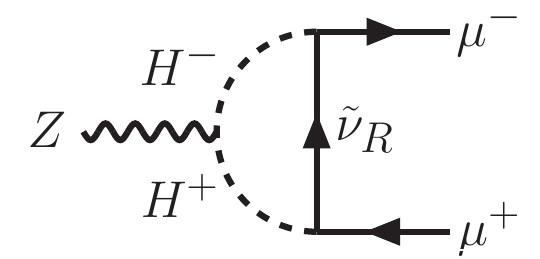} \qquad\quad
	\includegraphics[scale=0.5]{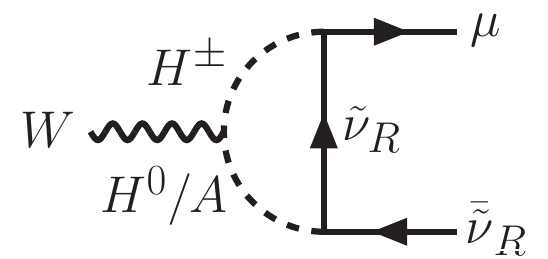}
	\caption{\label{fig.LFU} The NP contributions to the di-lepton decays $Z\to \mu^+\mu^-$ (left) and $W\to \mu \bar{\nu}$ (right).}
\end{figure}

For the $Z$-boson decays, the LFU ratio $R_{\mu \ell}^Z$ can be parameterized as
\begin{equation}\label{eq.ZLFU}
R_{\mu \ell}^Z \equiv \frac{\Gamma(Z\to \mu^+\mu^-)}{\Gamma(Z\to \ell^+\ell^-)} = R_{\mu \ell}^{Z, \rm{SM}} \left[ 1+\frac{2\,{\rm{Re}}\left(g_{V,Z}^{\rm{SM}} \cdot g_{V,Z}^{\mu,\rm{NP}\ast} + g_{A,Z}^{\rm{SM}} \cdot g_{A,Z}^{\mu,\rm{NP}\ast} \right)}{\left|g_{V,Z}^{\rm{SM}} \right|^2 + \left|g_{A,Z}^{\rm{SM}} \right|^2} \right],
\end{equation}
in the vanishing lepton mass limit. Here $R_{\mu \ell}^{Z, \rm{SM}}$ is the SM contribution and the SM couplings are given by $g_{V,Z}^{\rm{SM}} = -1/2 + 2 s_W^2$ and $g_{A,Z}^{\rm{SM}} = -1/2$, with $s^2_W\equiv \sin^2 \theta_W \simeq 0.23$~\cite{ParticleDataGroup:2020ssz}. It should be noted that the NP contribution to the electron/tauon mode is absent in view of the flavor-specific Yukawa structure given by Eq.~\eqref{Yukawa}. Given that the charged Higgs only couples to the left-handed muon (cf. Eq.~\eqref{muonphilic}) while the neutral scalars do not interact with the muon (cf. Eq.~\eqref{Lag2}), the NP contribution to the LFU ratio comes solely from the $H^+$-mediated loop diagram, as shown by the left Feynman diagram in Fig.~\ref{fig.LFU}. Explicitly, we arrive at
\begin{equation}
g_{V,Z}^{\mu,\rm{NP}} = g_{A,Z}^{\mu,\rm{NP}} = - \frac{\kappa_\nu^2 m_Z^2\, c_{2W}}{576\pi^2m_S^2},
\end{equation}
where $m_Z$ is the $Z$-boson mass and $c_{2W} = 1-2s_W^2$. Additionally, it should be mentioned that, in the quasi-degenerate limit for the Higgs mass spectrum as given by Eq.~\eqref{eq.Hmass}, the NP contributions to the decay $Z\to \mu^+\mu^-$ from the two neutral Higgs bosons $H$ and $A$ cancel out to a large extent, leaving therefore the dominant NP effect from the left Feynman diagram shown in Fig.~\ref{fig.LFU}. 

For the $W$-boson decays, the NP effect comes from the right Feynman diagram shown in Fig.~\ref{fig.LFU}. The resulting expression for the LFU ratio $R_{\mu \ell}^W$ in the vanishing lepton mass limit can be analogously obtained by replacing $Z$ with $W$ in Eq.~\eqref{eq.ZLFU}. The corresponding effective couplings are now given by $g_{V,W}^{\rm{SM}} = g_{A,W}^{\rm{SM}} =1/2$, and 
\begin{equation}
g_{V,W}^{\mu,\rm{NP}} = g_{A,W}^{\mu,\rm{NP}} = \frac{\kappa_\nu^2 m_W^2}{576\pi^2 m_S^2}.
\end{equation}
Note that, in deriving the above equation, we have made use of the quasi-degenerate Higgs mass spectrum given by Eq.~\eqref{eq.Hmass}. 

Then, by comparing the theoretical predictions with the experimental data~\cite{ParticleDataGroup:2020ssz,ATLAS:2020xea}
\begin{align}
	R_{\mu e}^{Z,\rm{exp}} &= 1.0001\pm 0.0024, \qquad R_{\mu e}^{W,\rm{exp}} = 0.993\pm 0.020,\\[0.2cm]
	R_{\mu \tau}^{Z,\rm{exp}} &= 0.9981\pm 0.0040, \qquad R_{\mu \tau}^{W,\rm{exp}} = 1.008\pm 0.013,
\end{align}
we can extract the bounds on the NP parameter space $(\kappa_\nu, m_S)$. To this end, we will take as constraints the experimental data within $1\sigma$ uncertainties.

\subsection{Constraints from the perturbative unitarity}
\label{PU}
In addition to the severe constraints from low-energy flavor physics, the theoretical considerations, such as the bounded-from-below limit on the scalar potential and the perturbative unitarity condition of the high-energy scattering amplitudes (see, e.g., Ref.~\cite{Branco:2011iw} and references therein), could render tight bounds on the NP parameter space as well. Here we will consider the vital requirement of perturbative unitarity for the Yukawa sector~\cite{Allwicher:2021rtd}.

Generically, the perturbative unitarity bounds can be derived in the so-called partial wave expansion approach~\cite{Jacob:1959at}. Explicitly, for the case of $2\to 2$ scattering processes in the high-energy massless limit, the partial waves $a_{fi}^J$ with fixed total angular momentum $J$ are defined by~\cite{Jacob:1959at}
\begin{equation}\label{eq:afiJ}
	a_{fi}^J = \frac{1}{32\pi} \int_{-1}^{1} d \cos \theta\,d_{\mu_i\mu_f}^J (\theta)\, \mathcal{T}_{fi} (\sqrt{s},\cos \theta),
\end{equation}
where $d_{\mu_i\mu_f}^J (\theta)$ are the small Wigner $d$-functions, with $\mu_i=\lambda_{i_1}-\lambda_{i_2}$ and $\mu_f=\lambda_{f_1}-\lambda_{f_2}$ characterizing the total helicities of the initial and final states respectively, and $\mathcal{T}_{fi} (\sqrt{s},\cos \theta)$ are the invariant scattering amplitudes, $(2\pi)^4\delta^{(4)}((p_{i_1}+p_{i_2})-(p_{f_1}+p_{f_2}))i\mathcal{T}_{fi} (\sqrt{s},\cos \theta)=\langle f|S-1|i \rangle$, that are related to the $S$ matrix by $S=1+i\mathcal{T}$. Here $\theta$ is the polar scattering angle in the center-of-mass frame, and $\sqrt{s}$ the center-of-mass energy. By focusing on the elastic channels, i.e., $|i \rangle =|f \rangle$, corresponding to a forward scattering with the same spin and internal variables in the initial- and final-state configurations, and restricting the sum over the intermediate states only to two-particle states, one can obtain from the unitarity condition of the $S$ matrix, $S^\dagger S=1$, the following reliable bounds on the tree-level partial-wave scattering matrices~\cite{Allwicher:2021rtd}:  
\begin{equation}\label{PUbound}
\left| a_{ii}^{J, \rm tree} \right| \le \frac{1}{2},
\end{equation}
which give a reasonable indication of the range of validity of the perturbative expansion. 

In order to extract the best perturbative unitarity bounds from Eq.~\eqref{PUbound}, we must then identify the optimal elastic channels. To this end, we have to know firstly the concrete expressions of the scattering amplitudes $\mathcal{T}_{fi} (\sqrt{s},\cos \theta)$, which depend on the definite Yukawa structure and scalar potential, as well as the underlying symmetry properties of the model considered. With $\mathcal{T}_{fi} (\sqrt{s},\cos \theta)$ at hand, it is then straightforward to obtain the perturbative unitarity bound for each entry of Eq.~\eqref{PUbound} by performing the convolution with the Wigner $d$-functions and the integration over the polar angle $\theta$ (cf. Eq.\eqref{eq:afiJ}), and then finding the eigenvalues of the partial-wave matrices $a_{fi}^J$. For the generic fermionic Yukawa interactions, due to the presence of different spin states in the scattering processes, this can be most efficiently achieved in the Jacob-Wick formalism~\cite{Jacob:1959at}. However, the traditional method for calculating $\mathcal{T}_{fi} (\sqrt{s},\cos \theta)$ relies on computing all the matrix entries, which becomes very involved and highly inefficient when the transition matrix has a very large dimension. Recently, it is noticed that the determination of perturbative unitarity bounds in this case can be simplified by decomposing each scattering amplitude with different $J$ into a Lorentz part that depends only on the spin and helicity of the fields involved and a group-theoretical part that depends only on their symmetry quantum numbers~\cite{Allwicher:2021rtd}. The only complication in the method is then attributed to the calculation of the symmetry factors, while the Lorentz parts are universal for different group structures~\cite{Allwicher:2021rtd}.

To obtain the perturbative unitarity bounds on the Yukawa parameters $\kappa_\nu$ and $\kappa_t$ of the $t\nu$2HDM, we employ here the results derived in Ref.~\cite{Allwicher:2021rtd}. For the lepton part, which is characterized by the SM gauge group $SU(2)_L \times U(1)_Y$, the most stringent bound comes from the $P$-wave amplitude, i.e., $J=1$, and imposes an upper limit on the muon-related coupling $\kappa_\nu$: 
\begin{equation}\label{upkappanu}
\kappa_\nu < \sqrt{4\pi \times \left(\sqrt{5}-1\right)} \approx 3.94.
\end{equation}
For the quark part, on the other hand, the constraints are quite different, since quarks carry an additional color quantum number under the gauge group $SU(3)_C$. As a consequence, the tightest constraint on the top-related coupling $\kappa_t$ stems from the $S$-wave amplitude with $J=0$, which leads to 
\begin{equation} \label{upkappat}
\kappa_t < \sqrt{8\pi/3} \approx 2.89,
\end{equation}
and hence a more stringent bound than on $\kappa_\nu$.

\subsection{Constraints from the LHC direct searches}
In the $t\nu$2HDM framework, as the quasi-degenerate mass regime in the alignment limit is considered, we can see that the $H^0$ and $A$ decay modes, $H^0/A\to AZ/ H^0 Z$, $H^0/A\to H^\pm W^\mp$ and $H^0\to AA, H^+ H^-$, are all forbidden, and the tree-level triple couplings $H^0/A-V-V$ (where $V$ denotes one of the gauge vector bosons $W/Z/\gamma$ and gluons) and $H^0/A-Z(h)-h$ are absent. This implies that for the heavy scalars concerned in this paper, their dominant decay modes are the top-quark and the neutrino pair, while the di-boson modes are suppressed by the loop factor and, more importantly, by the mass ratio $m_t/m_S$~\cite{Djouadi:2005gi}. Therefore, the decay width of the neutral scalars is approximately given by
\begin{align}\label{Swidth}
	\Gamma_{S} &\approx \Gamma (S\to t\bar t)+\Gamma (S\to\nu \bar\nu)
	\nonumber \\[0.2cm]
	&=\frac{m_S}{16\pi}\left[3\kappa_t^2\left(1-\frac{4m_t^2}{m_S^2}\right)^{n_{S}}+\kappa_\nu^2\right],
\end{align}
where $n_S=3/2$ for $S=H^0$ and $n_S=1/2$ for $S=A$, respectively.  

Currently, the LHC direct searches for the neutral scalar productions have been performed by the ATLAS with a center-of-mass energy $\sqrt{s}=8\,\mathrm{TeV}$~\cite{ATLAS:2017snw} and the CMS collaboration with $\sqrt{s}=13\,\mathrm{TeV}$~\cite{CMS:2019pzc} in the channel $pp\to S\to t\bar t$. In particular, the CMS results set model-independent constraints on the coupling modifiers $g_{S\bar t t}$ between the scalar $S$ and the top quark:
\begin{align}
	\mathcal{L}_{\bar t t S}=-g_{H^0 \bar t t} \frac{m_t}{v}\bar t t H^0+i g_{A \bar t t}\frac{m_t}{v}\bar t \gamma_5 t A.
\end{align}
The exclusion limits on $g_{S\bar t t}$ can then be translated into the allowed regions of the $t\nu$2HDM free parameters $\kappa_{t,\nu}$ and $m_S$. To this end, we must notice that the exclusion limits set by the CMS collaboration are obtained by assuming a fixed decay width $\Gamma_S$ with $\Gamma_S/m_S=[0.5,25]\%$. However, as can be inferred from Eq.~\eqref{Swidth}, for $\kappa_{t,\nu}\sim \mathcal{O}(1)$ and $m_S\gtrsim 500\,\mathrm{GeV}$, a ratio of $\Gamma_S/m_S\gtrsim (4-5)\%$ is obtained. As a consequence, we will only apply the two benchmark points, $\Gamma_S/m_S=10\%$ and $\Gamma_S/m_S=25\%$, selected in Ref.~\cite{CMS:2019pzc} to get a rough constraint on $\kappa_{t,\nu}$ for a fixed scalar  mass.

More significant constraints on the model parameters come from the LHC direct searches for the charged Higgs performed during the past few years. Both the ATLAS and CMS collaborations have covered several decay channels of the charged Higgs, which are dominated by the $\tau\nu$~\cite{ATLAS:2018gfm,CMS:2019bfg} and $tb$~\cite{ATLAS:2021upq,CMS:2020imj} final states. Recently, it is noticed in Ref.~\cite{Benbrik:2021wyl} that the $\mu \nu$ final state can also be an excellent complementary discovery channel of the charged Higgs. However, a comprehensive search for such a channel at the LHC is not available yet, and thus there is no any significant bound on the NP parameter space from the decay. Specific to the $t\nu$2HDM framework,  the decay modes of the charged Higgs are dominated by the $tb$ and $\mu\nu$ final states, and the $\tau\nu$ mode is suppressed under the flavor-specific Yukawa structure of Eq.~\eqref{Yukawa}, with the total decay width given approximately by 
\begin{align}\label{chargedwidth}
\Gamma_{H^+}&\approx \Gamma (H^+\to t\bar b)+\Gamma (H^+\to \mu^+ \nu)
\nonumber \\[0.2cm]
&=\frac{m_S}{16\pi}  \left[3\kappa_t^2|V_{tb}|^2\left(1-\frac{m_t^2}{m_S^2}\right)^2+ \kappa_\nu^2 \right]
\end{align}
where we have neglected the bottom-quark and muon masses. 

To obtain the viable parameter space of $(\kappa_{t},\kappa_\nu,m_S)$, we will apply the latest results from ATLAS~\cite{ATLAS:2021upq} and CMS~\cite{CMS:2020imj} with $\sqrt{s}=13\,\mathrm{TeV}$, where the model-independent exclusion limits on the $tb$-associated production cross section, $\sigma(pp\to H^\pm t b)$, times the branching fraction, $\mathcal{B}(H^\pm \to tb)$, are obtained for the charged-Higgs mass at $[0.2,2]\,\mathrm{TeV}$ and $[0.2,3]\,\mathrm{TeV}$ respectively, although the constraints from the CMS results are weaker than from the ATLAS searches. For the theoretical prediction in the $t\nu$2HDM framework, we have calculated the cross section $\sigma(pp\to H^\pm t b)$ using the computer program \texttt{MadGraph5\_{\small aMC}@NLO}~\cite{Alwall:2014hca},  with the charged-Higgs decay width given by Eq.~\eqref{chargedwidth}.  

\section{Mitigation of the $\boldsymbol{H_0}$ tension via the $\boldsymbol{R_{K^{(\ast)}}}$ resolution}
\label{contri}

\subsection{Two-parameter resolution of the $R_{K^{(\ast)}}$ anomalies}
\begin{figure}[t]
	\centering
	\subfigure[]{\label{fig.box}
	\begin{minipage}[htbp]{0.27\textwidth}
			\includegraphics[scale=0.45]{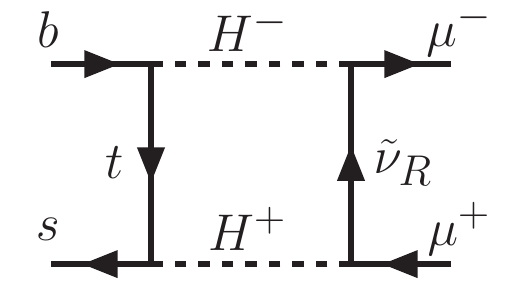}
	\end{minipage}} \hspace{0.2cm}
	\subfigure[]{\label{fig.Zpen}
		\begin{minipage}[htbp]{0.65\textwidth}
			\includegraphics[scale=0.40]{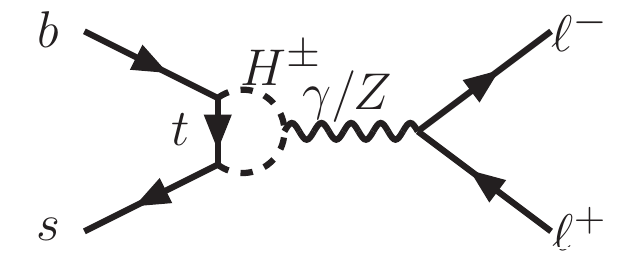}\qquad
			\includegraphics[scale=0.40]{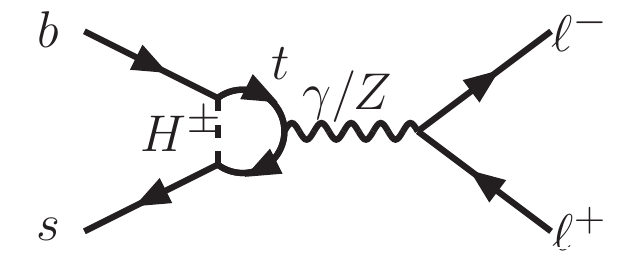}\\[0.5cm]
			\includegraphics[scale=0.40]{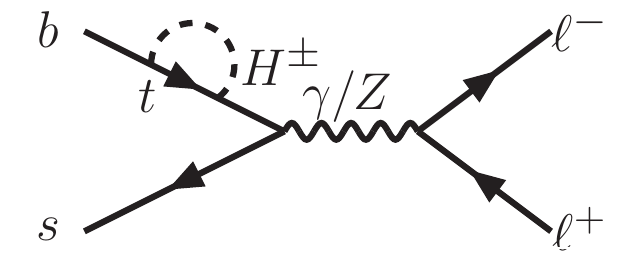}\qquad
			\includegraphics[scale=0.40]{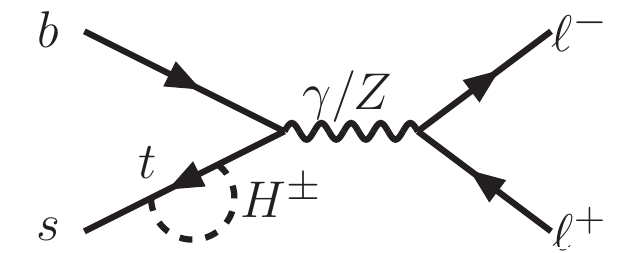}
	\end{minipage}}
	\caption{Fig.~(a): The NP box diagram contributing to the $b \to s \mu^+ \mu^-$ transition, where only one-flavor right-handed neutrino $\tilde{\nu}_{R}$ participates non-negligibly in the loop. Fig.~(b): The $\gamma/Z$-mediated NP penguin diagrams contributing to the $b\to s\ell^+\ell^-$ transition.} \label{RKD}
\end{figure}

The right-handed neutrino $\tilde{\nu}_R$ with its interaction specified by Eq.~\eqref{muonphilic} contributes to the $b\to s\ell^+\ell^-$ process mainly via the box diagram shown in Fig.~\ref{fig.box}. Its contribution can be described by the effective weak Hamiltonian
\begin{equation}
	\mathcal{H}_{\rm eff}^{\rm{NP}} = -\frac{  G_F}{\sqrt{2}} \frac{\alpha_e}{ \pi} V_{tb} V_{ts}^\ast \left( C_9  \mathcal{O}_{9} +C_{10} \mathcal{O}_{10}  \right) + \text{h.c.},
\end{equation}
where $\alpha_e=e^2/(4\pi)$ is the fine-structure constant, and the two effective operators are defined, respectively, as
\begin{align}
	\mathcal{O}_9 \equiv (\bar{s} \gamma_\mu P_L b) (\bar{\ell} \gamma^\mu \ell), \qquad \mathcal{O}_{10} \equiv (\bar{s} \gamma_\mu P_L b) (\bar{\ell} \gamma^\mu \gamma_5 \ell),
\end{align}
with the corresponding LFU-violating Wilson coefficients $C_{9\mu}^{\mathrm{NP}}$ and $C_{10\mu}^{\mathrm{NP}}$ given by
\begin{align} \label{C9C10}
	C_{9\mu}^{\mathrm{NP}} = -C_{10\mu}^{\mathrm{NP}} = \frac{- v^4 \left| \kappa_t \right|^2 \left| \kappa_\nu \right|^2}{64 s_W^2 m_W^2 m_{S}^2}  \frac{1- z_t +z_t \ln z_t}{\left( 1- z_t \right)^2}.
\end{align}
Note that we have neglected the neutrino mass in the above formula, and our result is consistent with that obtained in Ref.~\cite{Li:2018rax} in the vanishing neutrino mass limit.

In addition to the LFU-violating contribution given by Eq.~\eqref{C9C10}, the flavor-specific Yukawa texture characterized by Eq.~\eqref{Yukawa} also renders a considerable LFU-conserving effect on the $b\to s \ell^+\ell^-$ transition via the $\gamma/Z$-penguin diagrams shown in Fig.~\ref{fig.Zpen}. However, the resulting contributions to $C_{9\ell}^\textrm{NP}$ from the $\gamma$- and $Z$-penguin diagrams are suppressed by the factors $\alpha_e$ and $1-4s^2_W$, respectively. As a result, the dominant LFU-conserving contribution comes from the axial-vector part of the $Z$-boson couplings to fermions in the $Z$-penguin diagrams, with the final result given by 
\begin{equation}\label{C10Z}
C_{10\ell,Z}^\textrm{NP} = \frac{\kappa_t^2 v^2}{16 s_W^2 m_W^2} \frac{z_t \left( 1-z_t +\ln z_t \right)}{(1-z_t)^2}.
\end{equation}

It can be seen from Eqs.~\eqref{C9C10} and \eqref{C10Z} that, for $\kappa_{t,\nu} \sim \mathcal{O}(1)$ and $m_S \sim \mathcal{O}(1)\,\mathrm{TeV}$, the LFU-violating coefficients $C_{9\mu}^{\mathrm{NP}} = -C_{10\mu}^{\mathrm{NP}}$ have numerically the same order of magnitude as the LFU-conserving one $C_{10\ell,Z}^\textrm{NP}$. Interestingly, this observation is also favored by the two-parameter fit for the $R_{K^{(\ast)}}$ resolution~\cite{Alguero:2021anc}, 
\begin{equation}\label{twoparamfit}
C_{9\mu}^{\rm{NP}} = - C_{10\mu}^{\rm{NP}} =-0.53 \pm 0.10 , \qquad C_{10\ell,Z}^{\rm{NP}} = -0.24\pm 0.20,
\end{equation}
obtained at the $1\sigma$ level. While a negative central value of the LFU-conserving coefficient $C_{10\ell,Z}^{\rm{NP}}$ is, by itself, not helpful for explaining the $R_{K^{(\ast)}}$ anomalies, it can change the direction of the LFU-violating coefficients $C_{9\mu}^{\mathrm{NP}} = -C_{10\mu}^{\mathrm{NP}}$ and, in particular, lift $C_{9\mu}^{\mathrm{NP}}(=-C_{10\mu}^{\mathrm{NP}})$ to a larger negative value compared to the one-parameter fits obtained in Refs.~\cite{Alok:2019ufo,Carvunis:2021jga,Angelescu:2021lln,Geng:2021nhg,Cornella:2021sby,Kriewald:2021hfc,Alguero:2021anc,Hurth:2021nsi,Li:2021toq,Altmannshofer:2021qrr,Alok:2022pjb}. Translated to the parameter space in the $t\nu$2HDM framework, this requires larger $\kappa_{t,\nu}$ and/or lighter scalar mass $m_S$ to explain the $R_{K^{(\ast)}}$ anomalies. 

\subsection{Potential correlation between $R_{K^{(\ast)}}$ and $H_0$} 
As can be inferred from the previous studies made in Refs.~\cite{Iguro:2018qzf,Li:2018rax,Crivellin:2019dun,DelleRose:2019ukt}, an $\mathcal{O}(1)$ $\kappa_\nu$ is in general required to explain the $R_{K^{(\ast)}}$ anomalies. Specific to the $t\nu$2HDM framework, such a large coupling will readily help the right-handed neutrino $\tilde \nu_R$ establish thermal equilibrium with the SM plasma via the Higgs doublet portal $H_2$. When the temperature $T$ drops below the muon mass, the effective four-fermion interaction governing the right-handed neutrino annihilation rate, $\Gamma_{2\tilde{\nu}\to 2\nu}\equiv \Gamma (\tilde{\nu}_{R} \bar{\tilde\nu}_{R} \to\nu_L \bar{\nu}_L)$, mediated by the neutral scalars will determine the decoupling temperature $T_{\tilde{\nu}, \rm dec}$ of $\tilde{\nu}_R$. Since $\tilde\nu_R$ is relativistic in the early Universe, its contribution to the Hubble expansion can be parameterized by a shift of the effective neutrino number~\cite{Steigman:2012ve,Abazajian:2019oqj,Luo:2020sho,Adshead:2020ekg,Luo:2020fdt}
\begin{equation}\label{NEFF}
	\Delta N_{\rm eff} = N_{\tilde{\nu}} \frac{g_{\tilde{\nu}}} {2}\left( \frac{10.75}{g^s_\ast (T_{\tilde{\nu}, \rm dec})} \right)^{4/3}.
\end{equation}
Here $N_{\tilde{\nu}}=1$ denotes the number of thermalized right-handed neutrino species, and $g_{\tilde{\nu}} =2$ takes into account the antiparticle state of the right-handed Dirac neutrino, while $g_{\tilde{\nu}}=1$ for the right-handed Majorana neutrino. The effective d.o.f for the SM entropy density, $g^s_\ast (T_{\tilde{\nu}, \rm dec})$, is evaluated at the decoupling temperature $T_{\tilde{\nu}, \rm dec}$ of $\tilde{\nu}_R$, which can be estimated via the instantaneous decoupling condition, $\Gamma_{2\tilde{\nu}\to 2\nu}\simeq H(T_{\tilde{\nu},\rm dec})$, with the Hubble expansion rate given at the radiation-dominated epoch by
\begin{equation}\label{Hubble}
	H (T) = \sqrt{\frac{4\pi^3 g_\ast^\rho(T)}{45M_{P}^2}}\,T^2,
\end{equation}
where the effective d.o.f for the energy density is taken approximately as $g_*\equiv g_\ast^\rho \approx g_\ast^s$, and the Planck mass is given by $M_{P} =1.22\times 10^{19}\,\mathrm{GeV}$. 
  
The annihilation rate of the process $\tilde{\nu}_{R} \bar{\tilde\nu}_{R} \to\nu_L \bar{\nu}_L$ has the structure
\begin{equation}\label{eq:annihilationrate}
\Gamma_{2\tilde{\nu}\to 2\nu}= \frac{g_{\tilde{\nu}}}{2} \langle\sigma_{2\tilde{\nu}\to 2\nu} |v_{\tilde{\nu}_{R}}-v_{\bar{\tilde\nu}_{R}}|\rangle n_{\tilde\nu},
\end{equation}
where $|v_{\tilde{\nu}_{R}}-v_{\bar{\tilde\nu}_{R}}|$ is the relative velocity of the two incoming particles, $g_{\tilde{\nu}}/2$ is introduced to signify the symmetry factor due to the indistinguishability of particle and antiparticle in the initial state, and $n_{\tilde{\nu}}$ is the thermal particle number density of $\tilde{\nu}_R$ that is given by
\begin{align}
n_{\tilde\nu}=\frac{3\zeta(3)}{4\pi^2}T^3.
\end{align}
Here one should note that the spin d.o.f of the right-handed neutrino $\tilde{\nu}_R$ equals to one for both the chiral Dirac and Majorana neutrinos. The thermal rate $\langle\sigma_{2\tilde{\nu}\to 2\nu} |v_{\tilde{\nu}_{R}}-v_{\bar{\tilde\nu}_{R}}|\rangle$ in Eq.~\eqref{eq:annihilationrate} is given by
\begin{align}\label{thermalrate}
\langle\sigma_{2\tilde{\nu}\to 2\nu} |v_{\tilde{\nu}_{R}}-v_{\bar{\tilde\nu}_{R}}|\rangle &\equiv \frac{\int dn^{\rm eq}_{\tilde{\nu}}(p_1) dn^{\rm eq}_{\tilde{\nu}} (p_2)\,\sigma_{2\tilde{\nu}\to 2\nu} |v_{\tilde{\nu}_{R}}-v_{\bar{\tilde\nu}_{R}}|}{\int dn^{\rm eq}_{\tilde{\nu}}(p_1)dn^{\rm eq}_{\tilde{\nu}}(p_2)}
\nonumber \\[0.2cm]
&=\frac{T}{32\pi^4 n_{\tilde\nu}^2}\int _{0}^\infty d\hat s\,\sigma_{2\tilde{\nu}\to 2\nu}\, \hat{s}^{3/2}\,K_1(\sqrt{\hat  s}/T),
\end{align}
where $K_1(x)$ is the modified Bessel function of order one, and the phase-space factor is defined by
\begin{align}
dn^{\rm eq}_{\tilde{\nu}}(p_i)=\frac{d^3p_i }{(2\pi)^3}  f^{\rm eq}_{\tilde{\nu}}(p_i).
\end{align} 
Within the $t\nu$2HDM framework, the annihilation cross section $\sigma_{2\tilde{\nu}\to 2\nu}$ is simply given by
\begin{align}
\sigma_{2\tilde{\nu}\to 2\nu}= \frac{\kappa_\nu^4 }{192\pi m_S^4}\hat{s},
\end{align}
where $\sqrt{\hat s}=E_{\rm cm}$ is the center-of-mass energy. Finally, we obtain the annihilation rate of the process $\tilde{\nu}_{R} \bar{\tilde\nu}_{R} \to\nu_L \bar{\nu}_L$, 
\begin{align}
	\Gamma_{2\tilde{\nu}\to 2\nu} =\frac{g_{\tilde{\nu}}}{2}\frac{\kappa_\nu^4}{6\zeta (3) \pi^3} \frac{T^5}{m_S^4},
\end{align}
which, together with the instantaneous decoupling condition $\Gamma_{2\tilde{\nu}\to 2\nu}\simeq H(T_{\tilde{\nu},\rm dec})$ and Eq.~\eqref{Hubble}, leads to the decoupling temperature,
\begin{align}\label{Tndec}
	\left(\frac{T_{\tilde{\nu},\rm dec}}{\mathrm{MeV}}\right)\simeq 4.25 \left(\frac{2}{g_{\tilde{\nu}}}\right)^{1/3} \left(\frac{g_*(T_{\tilde{\nu},\rm dec})}{10.75}\right)^{1/6}\left(\frac{3}{\kappa_{\nu}}\right)^{4/3}\left(\frac{m_S}{500\,\mathrm{GeV}}\right)^{4/3}. 
\end{align}
It should be mentioned that, to obtain the analytic thermal rate as given by Eq.~\eqref{thermalrate}, we have used the Boltzmann distribution $f^{\rm eq}_{\tilde{\nu}}=e^{-E/T}$. Since the dependence of the d.o.f $g_*^s(T)$ on the decoupling temperature $T_{\tilde{\nu}, \rm dec}$ is weak below the muon mass scale $T<m_\mu$~\cite{Husdal:2016haj}, the effective neutrino number shift $\Delta N_{\rm eff}$ will also have a weak dependence on $T_{\tilde{\nu}, \rm dec}$. Therefore, the approximation of adopting the Boltzmann distribution is sufficient to estimate the scale of $T_{\tilde{\nu}, \rm dec}$ from Eq.~\eqref{Tndec}.

From Eq.~\eqref{Tndec}, one can see that the decoupling temperature $T_{\tilde{\nu}, \rm dec}$ will be solely determined by the free parameters $\kappa_\nu$ and $m_S$ after inserting the effective d.o.f $g_*(T)$ as a function of the temperature~\cite{Husdal:2016haj}. This in turn implies that the effective d.o.f $g_*^s(T_{\tilde{\nu},\rm dec})$ present in Eq.~\eqref{NEFF} and hence the effective neutrino number shift $\Delta N_{\rm eff}$ are also determined by the two parameters $\kappa_\nu$ and $m_S$. On the other hand, given that the parameter $\kappa_t$ is severely constrained by the low-energy flavor physics (especially by the mass differences $\Delta M_q$), we know that $\kappa_{\nu}$ becomes the key parameter for the $R_{K^{(\ast)}}$ resolution. Therefore, one can expect that there must exist a potential correlation between the $R_{K^{(\ast)}}$ resolution and the mitigation of the $H_0$ tension achieved via the effective neutrino number shift given by Eq.~\eqref{NEFF}, within the $t\nu$2HDM framework proposed here.

\section{Numerical results and discussions}
\label{num}

\subsection{Viable parameter space for the $R_{K^{(\ast)}}$ resolution}
\begin{figure}[t]
	\centering
	\includegraphics[scale=0.41]{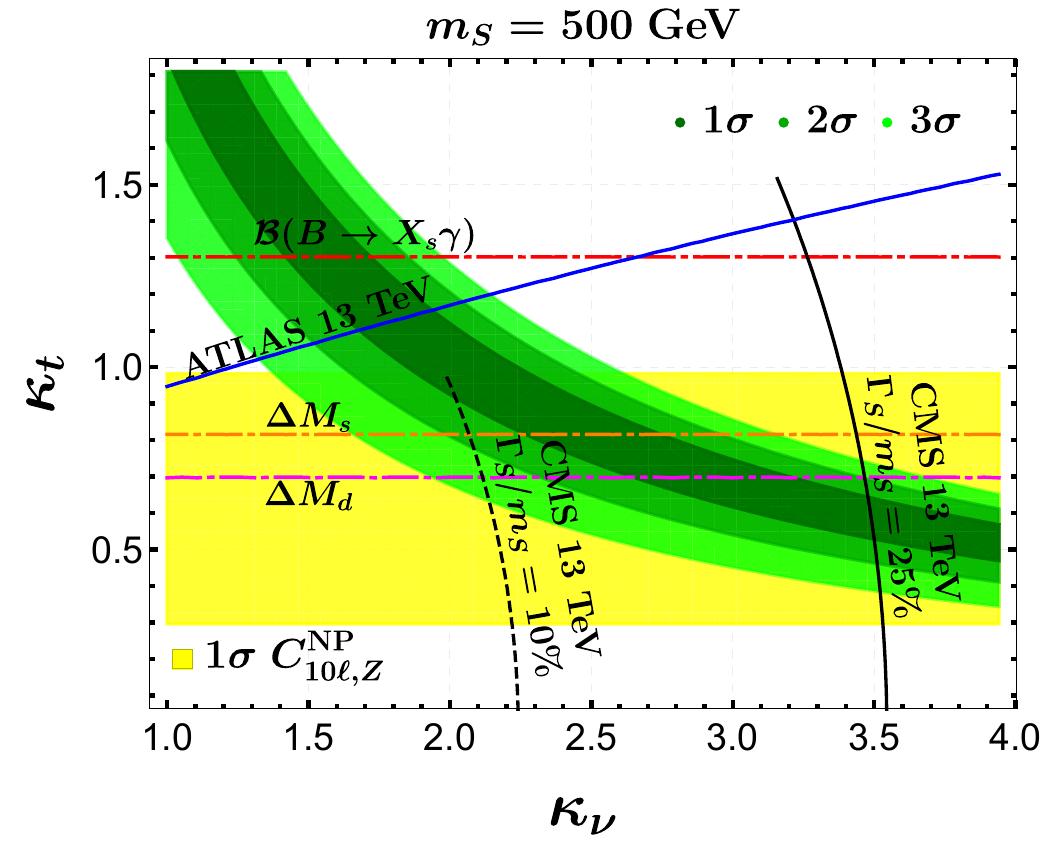}\quad
	\includegraphics[scale=0.41]{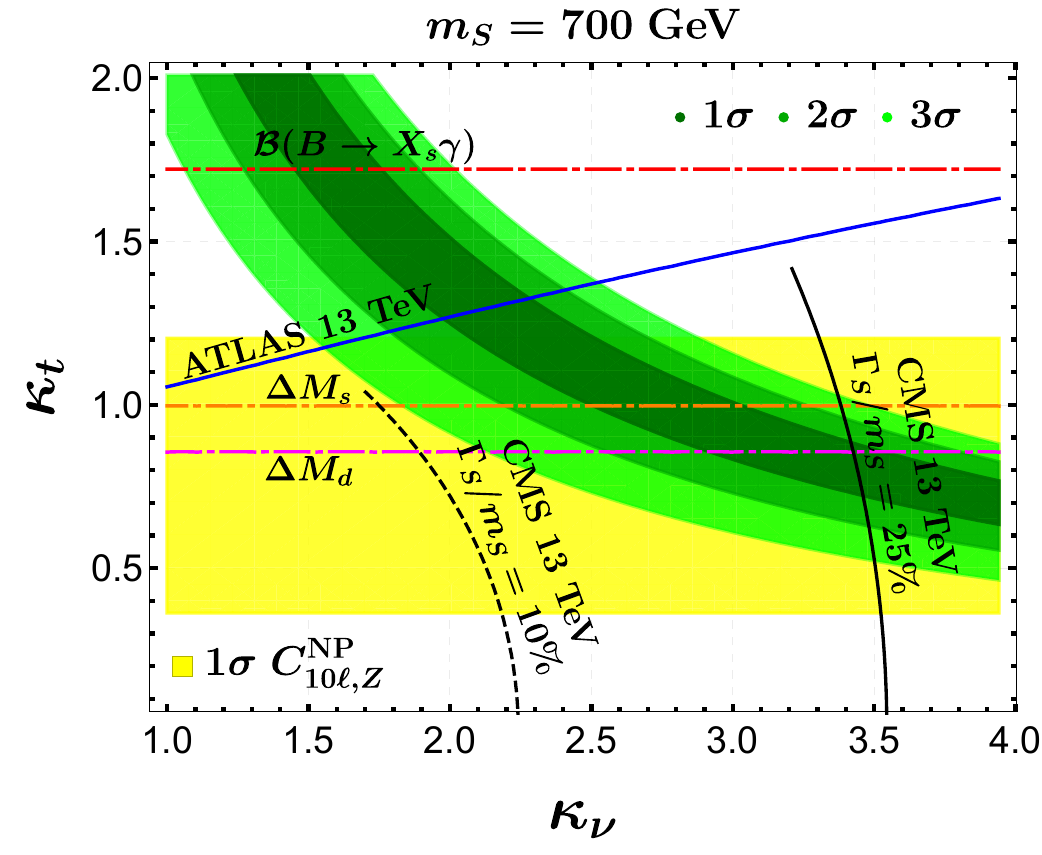}\\[0.4cm]
	\includegraphics[scale=0.41]{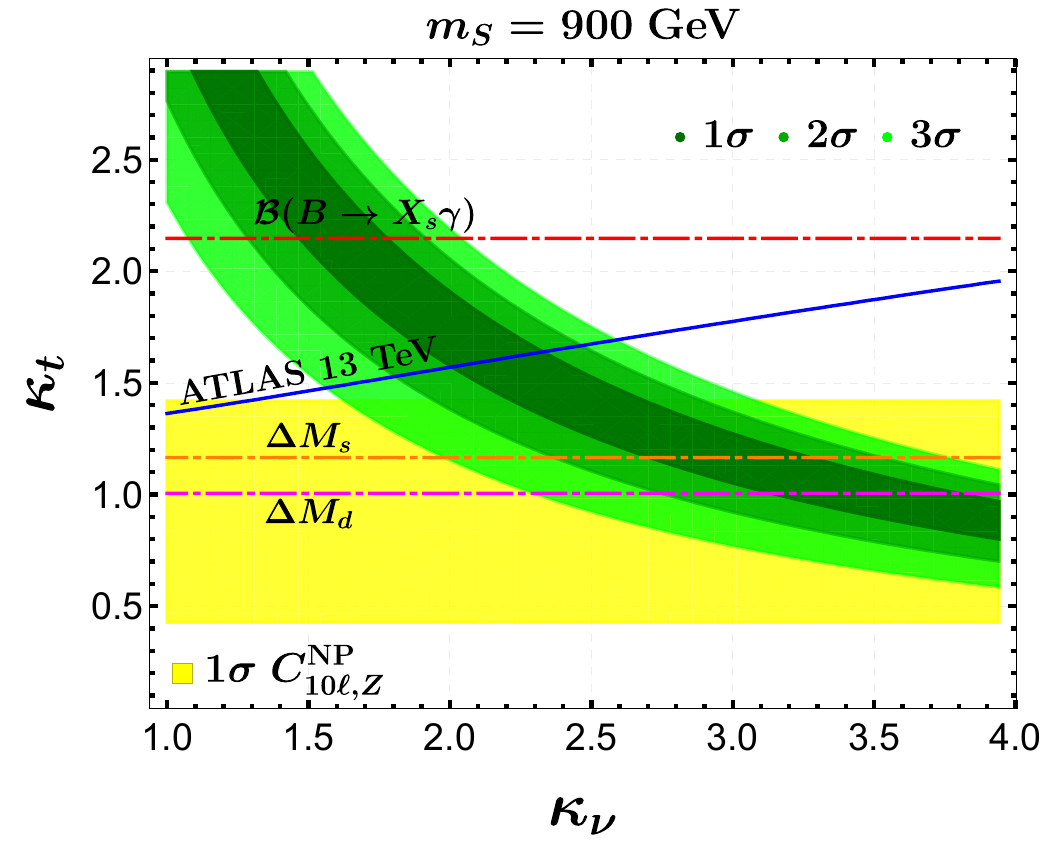}\quad
	\includegraphics[scale=0.41]{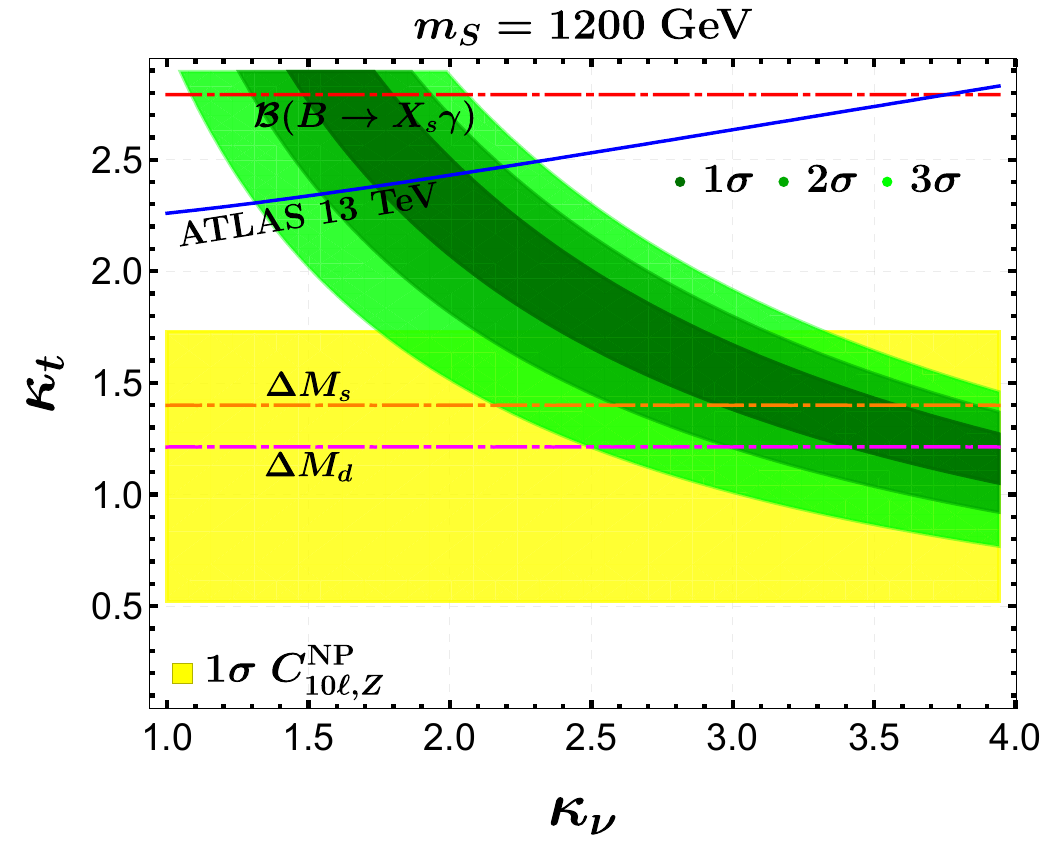}
	\caption{The viable parameter space in the $(\kappa_\nu,\kappa_t)$ plane for the $R_{K^{(\ast)}}$ resolution, with the Higgs mass fixed at $m_S = 500$, $700$, $900$ and $1200\,{\mathrm{GeV}}$, respectively. The green and yellow bands represent the regions allowed by the two-parameter fits~\cite{Alguero:2021anc} in the direction of $C_{9\mu}^{\rm{NP}} = -C_{10\mu}^{\rm{NP}}$ and $C_{10\ell,Z}^{\rm NP}$, as given by Eq.~\eqref{twoparamfit}. We have also taken into account all the relevant constraints discussed in the last two sections; see text for further details.} \label{fig.ktkn}
\end{figure}

Let us begin with the exploration of the NP parameter space allowed by the $R_{K^{(\ast)}}$ anomalies. By fixing the quasi-degenerate Higgs mass at $m_S = 500$, $700$, $900$ and $1200\,{\mathrm{GeV}}$ respectively, we show in Fig.~\ref{fig.ktkn} the viable parameter regions in the $(\kappa_\nu,\kappa_t)$ plane, under the perturbative unitarity bounds given by Eqs.~\eqref{upkappanu} and \eqref{upkappat}. We have also taken into account all the relevant phenomenological constraints discussed in Sec.~\ref{pheno}. Explicitly, the regions above the various curves in Fig.~\ref{fig.ktkn} are already excluded by the branching ratio $\mathcal{B}(\bar{B}\to X_s\gamma)$ (red), the mass differences $\Delta M_s$ (orange) and $\Delta M_d$ (magenta), as well as the direct searches for the charged Higgs from ATLAS with $13\,\mathrm{TeV}$ (blue). In the upper two plots, we also show the correlation between $\kappa_t$ and $\kappa_\nu$ inferred from the CMS direct searches for the neutral scalars, with two benchmark points of the decay width over mass ratio, $\Gamma_S/m_S = 10\%$ (black dashed) and $\Gamma_S/m_S = 25\%$ (black solid). As the LFU ratios of the di-lepton decays of $Z/W$ gauge bosons do not impose any further significant constraints in the $(\kappa_\nu,\kappa_t)$ plane under the perturbative unitarity bounds, they are not displayed in Fig.~\ref{fig.ktkn}. Finally, the bands colored from the dark to the light green represent the regions allowed by the $R_{K^{(\ast)}}$ resolution in the direction of $C_{9\mu}^{\rm{NP}} = -C_{10\mu}^{\rm{NP}}$ at the $1\textendash3\sigma$, while the band in yellow denotes the $1\sigma$ region of $C_{10\ell,Z}^{\rm NP}$, as given in Eq.~\eqref{twoparamfit}. 

From Fig.~\ref{fig.ktkn}, it can be readily seen that the most stringent bound in the quark sector comes from the mass differences $\Delta M_q$ and, in particular, from $\Delta M_d$, which in turn requires $\kappa_\nu\simeq 3$ for the $R_{K^{(\ast)}}$ resolution. However, as mentioned already in Sec.~\ref{BBmixing}, the $\Delta M_q$ constraints may not be so conclusive due to the large theoretical uncertainties. It should also be pointed out that, if the $\Delta M_q$ discrepancies observed in Sec.~\ref{BBmixing} were confirmed in the future, we have to resort to some extra NP sources beyond the minimal $t\nu$2HDM setup considered here. In such a special case, the constraints from $\Delta M_s$ (orange) and $\Delta M_d$ (magenta) may become irrelevant. On the other hand, for $m_S = 500$ and $m_S = 700\,{\mathrm{GeV}}$, the two black curves inferred from the CMS direct searches for the neutral scalars should be interpreted as the maximal values of $\kappa_t$ under the reference values of $\kappa_\nu$. For instance, with $\kappa_\nu\approx2$, $\kappa_t>0.97$ will be excluded by the limits set by the CMS direct searches for the process $pp\to S\to t \bar t$~\cite{CMS:2019pzc}. It can also be seen that, for the benchmark point $\Gamma_S/m_S = 10\%$ and $m_S=500\,\mathrm{GeV}$, the constraint on $\kappa_t$ from the CMS neutral-scalar searches is tighter than from the charged-Higgs bound set by the ATLAS collaboration with $13\,\mathrm{TeV}$~\cite{ATLAS:2021upq}, while for $\Gamma_S/m_S$ reaching up $25\%$ and $m_S=700\,\mathrm{GeV}$, the upper limit on $\kappa_t$ is still determined by the CMS neutral-scalar searches. However, we must note that the CMS constraints are no longer applicable for $m_S>750\,\mathrm{GeV}$~\cite{CMS:2019pzc}.
 
In the next subsection, we will see that a large muon-related coupling $\kappa_\nu$ as required by the $R_{K^{(\ast)}}$ resolution is necessary for generating a significant contribution to the $\Delta N_{\rm{eff}}$ shift. In this respect, we conclude that the $t\nu$2HDM framework provides us with an opportunity to correlate the $R_{K^{(\ast)}}$ resolution with the mitigation of the $H_0$ tension.

\subsection{Favored $\Delta N_{\rm eff}$ shift for the $H_0$ tension}

\begin{figure}[t]
	\centering
	\subfigure[]{\label{fig.RKH0500}
		\begin{minipage}[b]{0.5\textwidth}
			\includegraphics[scale=0.5]{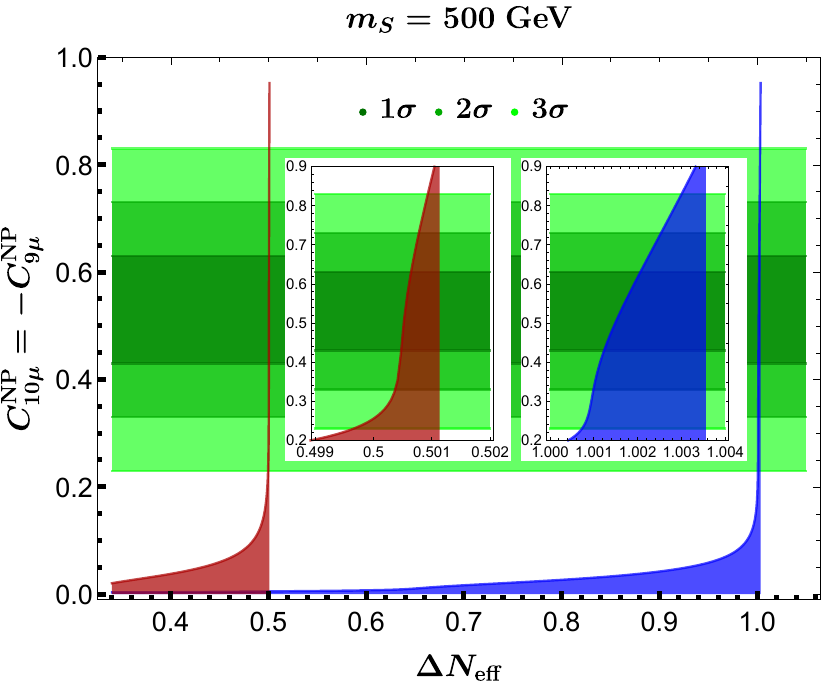}
	\end{minipage}} \hspace{-0.5cm}
	\subfigure[]{\label{fig.RKH01000}
		\begin{minipage}[b]{0.5\textwidth}
			\includegraphics[scale=0.5]{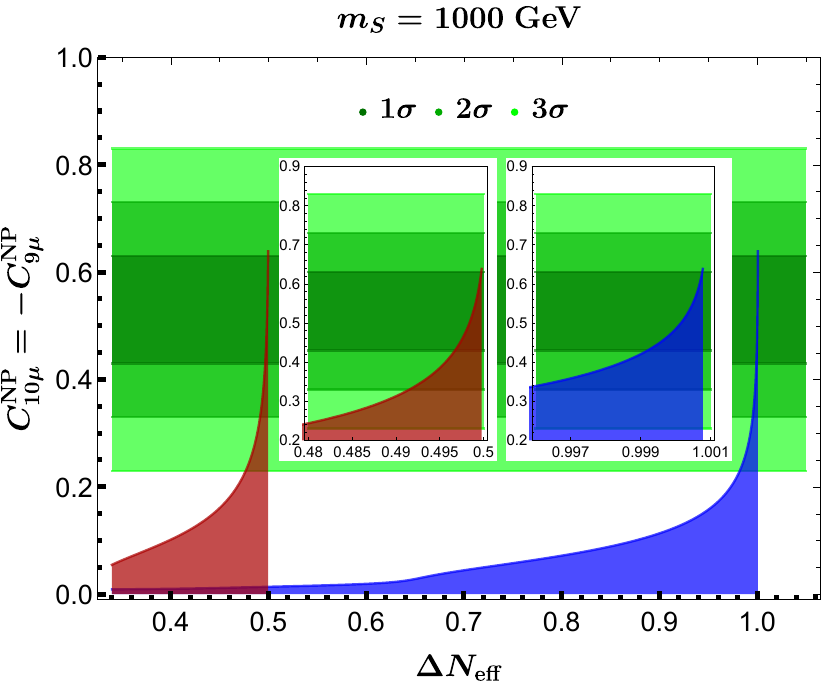}
	\end{minipage}}
	\caption{The induced ranges of $C_{10\mu}^{\rm{NP}}=-C_{9\mu}^{\rm{NP}}$ for given values of $\Delta N_{\rm{eff}}$, with the parameter $\kappa_t$ varied from zero up to the values allowed by the $\Delta M_q$ constraints, and the scalar mass fixed at $m_S = 500\,{\mathrm{GeV}}$ (Fig.~(a)) and $m_S = 1000\,{\mathrm{GeV}}$ (Fig.~(b)), respectively. The colored regions in blue and red correspond to the Dirac and the Majorana nature of the right-handed neutrino, respectively. The horizontal bands colored in green correspond to the global-fit results of $C_{10\mu}^{\rm{NP}}(=-C_{9\mu}^{\rm{NP}})$ at the $1\textendash 3\sigma$ level, as given in Eq.~\eqref{twoparamfit}.} \label{RKH0}
\end{figure}

To visualize the potential correlation between the $R_{K^{(\ast)}}$ resolution and the mitigation of the $H_0$ tension achieved via the $\Delta N_{\rm{eff}}$ shift, we start with Eq.~\eqref{NEFF}, where the effective d.o.f $g_*^s(T_{\tilde{\nu},\rm dec})$ is solely determined by the free parameters $\kappa_\nu$ and $m_S$ within our approximation. The LFU-violating Wilson coefficients $C_{9\mu}^{\rm{NP}} = -C_{10\mu}^{\rm{NP}}$ in Eq.~\eqref{C9C10} can then be expressed in terms of $\kappa_t$, $m_S$ and $\Delta N_{\rm{eff}}$. By fixing the scalar mass $m_S$ and varying the parameter $\kappa_t$ from zero up to the values allowed by the $\Delta M_q$ constraints, we can finally obtain the numerical dependence of $C_{9\mu}^{\rm{NP}} = -C_{10\mu}^{\rm{NP}}$ on $\Delta N_{\rm{eff}}$, which are shown in Figs.~\ref{fig.RKH0500} and \ref{fig.RKH01000} for $m_S = 500$ and $m_S = 1000\,{\mathrm{GeV}}$, respectively. The horizontal bands colored from the dark to the light green correspond to the global-fit results of $C_{9\mu}^{\rm{NP}} = -C_{10\mu}^{\rm{NP}}$ at the $1\textendash 3\sigma$ level, as given in Eq.~\eqref{twoparamfit}. The blue (red) region corresponds to the viable parameter space in the Dirac (Majorana) neutrino case, in which the peak of $C_{9\mu}^{\rm{NP}} = -C_{10\mu}^{\rm{NP}}$ corresponds to the upper limit on $\kappa_{\nu}$ as required by the perturbative unitarity bound (see Eq.~\eqref{upkappanu}).

From Fig.~\ref{RKH0}, it can be clearly seen that a resolution of the $R_{K^{(\ast)}}$ anomalies at the $1\sigma$ level requires a shift $\Delta N_{\rm{eff}} \simeq 1.0$ for a one-flavor right-handed Dirac neutrino and $\Delta N_{\rm{eff}} \simeq 0.5$ for a one-flavor right-handed Majorana neutrino, and any large or moderate deviation from the benchmark points of $\Delta N_{\rm eff}$, although being able to ease the $H_0$ tension, cannot resolve the $R_{K^{(\ast)}}$ anomalies at the same time. In both cases, after fixing the Higgs mass, a large effective $\Delta N_{\rm eff}$ is always required by an approximately $\kappa_\nu \simeq 3$ that coincides almost with the perturbative unitarity limit, and varying the value of $\kappa_t$ only influences the Wilson coefficients $C_{9\mu}^{\rm{NP}} = -C_{10\mu}^{\rm{NP}}$. Besides, by comparing the two figures, we can find that increasing the Higgs mass will enlarge the viable space of $\Delta N_{\rm eff}$ and, at the same time, shrink the range of $C_{9\mu}^{\rm{NP}} = -C_{10\mu}^{\rm{NP}}$. This indicates a preference of a lighter Higgs to address the $R_{K^{(\ast)}}$ anomalies while easing the $H_0$ tension. In addition, such a difference between Dirac and Majorana neutrinos is expected due to the different spinor natures of the neutrinos involved, i.e., the Weyl spinor for the former and the Majorana spinor for the latter case. In terms of the favored $\Delta N_{\rm{eff}}$ shifts inferred from Eqs.~\eqref{R19noBBN}--\eqref{newBBN}, we can then conclude that the eV-scale Majorana neutrino embedded in the $t\nu$2HDM framework is able to address the $R_{K^{(\ast)}}$ anomalies and, at the same time, ease the $H_0$ tension, while the case with one-flavor Dirac neutrino generates a too large $\Delta N_{\rm{eff}}$ shift.

Finally, it can be demonstrated that, if more than one neutrino contributes to the  $R_{K^{(\ast)}}$ anomalies via the box diagram shown in Fig.~\ref{fig.box}, the resulting $\Delta N_{\rm{eff}}$ shift would be unacceptably large. As an example, let us consider the case where there are two right-handed neutrinos having significant couplings to the muon lepton in Eq.~\eqref{Lag2}. Then, we must sum over the two flavors of $\nu_R$ in Eq.~\eqref{C9C10}, i.e., $|\kappa_\nu|^2\to |\kappa_{\nu,1}|^2+|\kappa_{\nu,2}|^2$. Assuming that $\kappa_{\nu,1}\sim \kappa_{\nu,2}$ and applying our previous finding $\kappa_\nu\simeq 3$ for the $R_{K^{(\ast)}}$ resolution as inferred from Fig.~\ref{fig.ktkn}, we can see that $\kappa_{\nu,1}\sim \kappa_{\nu,2}\sim \mathcal{O}(3/\sqrt{2})$ are required in this case. This means that the muon-related couplings $\kappa_\nu$ can be reduced by a factor of $1/\sqrt{2}$ in explaining the $R_{K^{(\ast)}}$ anomalies with two right-handed neutrinos. However, such a parameter reduction cannot cause any significant lift of the decoupling temperature and, more importantly, any significant increase of the effective d.o.f $g_*(T_{\tilde{\nu},\rm dec})$, in Eq.~\eqref{Tndec}. This can be understood by the fact that enhancing $T_{\tilde{\nu},\rm dec}$ by a factor of $2^{2/3}$ can only increase $g_*(T_{\tilde{\nu},\rm dec})$ by about $10\%$~\cite{Husdal:2016haj}. Then, one can see from Eq.~\eqref{NEFF} that the $\Delta N_{\rm{eff}}$ shift is basically determined by the number of thermalized right-handed neutrino species. As a consequence, the correlation shown in Fig.~\ref{RKH0} indicates that the $R_{K^{(\ast)}}$ resolution with more than one thermalized right-handed neutrino would entail a large $\Delta N_{\rm{eff}}$ shift beyond that favored by Eqs.~\eqref{R19noBBN}--\eqref{newBBN}. This is the reason why we have introduced only the one-flavor thermalized right-handed neutrino $\tilde{\nu}_R$ into the early Universe within our framework.

\section{Conclusion}
\label{con}

The latest updated measurements from the LHCb~\cite{LHCb:2021trn} and SH0ES~\cite{Riess:2021jrx} collaborations have respectively strengthened the deviations of the LFU ratio $R_{K}$ in rare semi-leptonic $B$-meson decays and the present-day $H_0$ parameter in the Universe. If confirmed with more precise experimental measurements and theoretical predictions, they could be tantalizing hints of NP beyond the SM. In this paper, we have constructed a simple flavor-specific 2HDM, dubbed the $t\nu$2HDM, where significant NP effects arise only from the one-flavor right-handed neutrino and the top quark. Such a framework is only characterized by the three free parameters $\kappa_t$, $\kappa_\nu$, and $m_S$, in the alignment limit with a quasi-degenerate Higgs mass spectrum. 

The $t\nu$2HDM can explain the long-standing $R_{K^{(\ast)}}$ anomalies via one eV-scale right-handed Majorana neutrino or one right-handed Dirac neutrino, under the most relevant constraints from the low-energy flavor physics, the perturbative unitarity condition, as well as the LHC direct searches. However, being different from the three-flavor right-handed neutrino scenarios considered in Refs.~\cite{Iguro:2018qzf,Li:2018rax,Crivellin:2019dun,DelleRose:2019ukt}, one of the intriguing predictions resulting from the parameter space for the $R_{K^{(*)}}$ resolution with such a one-flavor scenario points toward a moderate shift of the effective neutrino number, $\Delta N_{\rm eff}=N_{\rm  eff}-N_{\rm eff}^{\rm SM}$, at the early BBN and late CMB epochs. It is then found that, while the $\Delta N_{\rm eff}$ shift predicted in the Dirac neutrino case is still at $\Delta N_{\rm eff}\simeq 1.0$ and hence disfavored by the CMB polarization measurements, the one induced in the Majorana case is $\Delta N_{\rm eff}\simeq 0.5$, which coincides with the ranges from Eqs.~\eqref{R19noBBN}--\eqref{newBBN} favored to ease the notorious $H_0$ tension~\cite{Seto:2021xua,Seto:2021tad,Matsumoto:2022tlr,Kawasaki:2022hvx,Burns:2022hkq}. There also exists a potential correlation between the $R_{K^{(\ast)}}$ anomalies and the $H_0$ tension achieved via the $\Delta N_{\rm eff}$ shift with the one-flavor eV-scale right-handed Majorana neutrino, and such a correlation can be tested in the future. 

As a conclusion, the $t\nu$2HDM provides an interesting link between the $R_{K^{(*)}}$ anomalies and the $H_0$ tension. In addition, a light right-handed Majorana neutrino embedded in the 2HDM infers a hierarchical Majorana neutrino pattern for the seesaw generation of the neutrino masses and, in particular, a nearly massless active neutrino. 

As a last comment, we give here some discussions about the direct searches of the right-handed neutrinos. These right-handed neutrinos, which are also called the heavy neutral leptons with masses above the eV scale, are often proposed to explain several puzzles of fundamental physics, first and foremost — the neutrino oscillations. These hypothetical particles can be of Majorana or of Dirac nature. The present generation of experiments usually focus on the following three aspects: neutrino masses, oscillation parameters, and neutrinoless double beta decay~\cite{Coloma:2022dng,Cirigliano:2022oqy}. Future precise measurements of these parameters can come from many kinds of experiments, such as the short-baseline experiments, the fixed-target experiments, the collider experiments, and so on. With the upcoming precision era of neutrino physics, these terrestrial experiments are expected to determine the exact mixing pattern and flavor structures of the heavy neutral leptons~\cite{Coloma:2022dng}. In addition, specific to the $t\nu$2HDM, the new interactions in the lepton sector can lead to the charged-Higgs decaying into the right-handed neutrinos, $H^+\to\mu^+\nu$. These right-handed neutrinos can be, therefore, searched for at the LHC in terms of the SM-like Yukawa interactions with the extended neutrinos. However, such kinds of processes have not been observed at the LHC by now, and only some phenomenological studies exist in the literature~\cite{Benbrik:2021wyl}. We expect that the right-handed neutrinos can be detected via the channel $H^+\to\mu^+\nu$ in the future experiments, and the free parameters related to the heavy neutral leptons can be determined by the forthcoming neutrino experiments.

\acknowledgments
We thank Biao-Feng Hou for providing us with the \texttt{MadGraph5\_{\small aMC}@NLO} calculation and helpful discussions. This work is supported by the National Natural Science Foundation of China under Grant Nos.~12135006, 12075097, 11675061 and 11775092, as well as by the Fundamental Research Funds for the Central Universities under Grant Nos.~CCNU20TS007, CCNU19TD012 and CCNU22LJ004.

\bibliographystyle{JHEP}
\bibliography{references}

\end{document}